\documentclass[twocolumn]{aastex631}
\usepackage{amsmath,amstext}
\usepackage[T1]{fontenc}
\usepackage{apjfonts} 
\usepackage{multirow}
\usepackage[figure,figure*]{hypcap}
\usepackage{booktabs,caption}
\usepackage{tablefootnote}
\usepackage{hyperref}
\usepackage{subcaption}
\usepackage{mwe}

\begin{document}

\title{The AstroSat UV Deep Field South II: A diverse set of Lyman-continuum leakers at $z\sim1$}

\author[0009-0003-8568-4850]{Soumil Maulick}
\affil{Inter-University Centre for Astronomy and Astrophysics, Ganeshkhind, Post Bag 4, Pune 411007, India}

\author[0000-0002-8768-9298]{Kanak Saha}
\affil{Inter-University Centre for Astronomy and Astrophysics, Ganeshkhind, Post Bag 4, Pune 411007, India}

\author[0000-0001-7016-5220]{Michael J. Rutkowski}
\affiliation{Minnesota State University-Mankato, Mankato, MN 56001, USA}



\begin{abstract}
We report the detection of 5 newly identifed Lyman-continuum (LyC) leaker candidates at redshifts 0.99-1.42 in the AstroSat UV Deep Field South F154W image. We derive physical properties of these galaxies using a combination of spectral-energy distribution fitting and information from publicly available spectra. The estimated escape fraction of these objects vary from 14-85\% after accounting for the IGM attenuation. With only about a dozen known leakers at these redshifts, these detections significantly raise the fraction of LyC leakers in this redshift range. High-resolution HST UV imaging reveals that a subset of the galaxies in our sample have blue star-forming structures that are likely associated with harder ionizing sources. We find tentative evidence that the LyC emission is spatially offset from the non-ionizing UV continuum centers of these galaxies. The integrated properties of these galaxies, such as the UV continuum slope, dust attenuation, stellar mass, and $[\text{O III}] \lambda 5007/ [\text{O II}] \lambda 3727$ ratios, make them atypical compared to known LyC leakers. The leakage of LyC photons from these systems presents a compelling challenge.  
\end{abstract}



\section{Introduction}
The Epoch of Reionization (EOR)  marks an important period in the cosmic history of the Universe when the neutral hydrogen in the intergalactic medium (IGM) in the universe transitioned from neutral to ionized \citep{Barkana01}. Recent results indicate that this process was completed between redshifts $z\sim5-6 $ \citep{Bosman22,Zhu24}. To understand the nature of reionization, it is necessary to account for the sources of ionizing photons. One of the promising candidates for these sources are thought to be star-forming galaxies (SFGs) \citep{Robertson15}. However, direct detection of Lyman-continuum (LyC) photons from these reionization era sources is infeasible because of the high optical depth of the IGM along the line-of-sight to these redshifts \citep{Inoue14}. 
\par Lower redshift analogs are then studied with the hope that they can provide some insights into the relevant properties of the reionization era sources. Substantial efforts have been devoted to identify LyC leakers at low redshift. These include studies that have reported non-detections (e.g., \cite{Siana07, Rutkowski16, Alavi20, Flury22a, Jung24}) and provided constraints that relate to LyC escape. At $z\sim 0.2-0.4$ multiple LyC leakers have now been detected using spectroscopic observations with the HST COS instrument \citep{Borthakur2014,Izotov2016,Izotov16b,Izotov18,Izotov18b,Izotov21,Wang19,Flury22a}. Between redshifts $\sim 1-1.5$, the UVIT telescope onboard AstroSat has detected LyC leakers in photometry \citep{Saha20,Dhiwar24,Maulick24}. Beyond redshifts of 2 the HST UV bands (F275W and F336W) have been successful in the photometric detection of LyC leakers; $z\sim 2-2.5$ \citep{Bian17,RiveraThorsen19,Smith24}; $z\sim 3-4$ \citep{Mostardi15,Vanzella16,Yuan21,Kerutt23}. Spectroscopic observations with VLT/VIMOS, MOSFIRE/Keck and GTC/OSIRIS have further revealed LyC leakers at $z\sim 3-4$ \citep{deBarros16,Vanzella18,Steidel18,MarquesChaves21,MarquesChaves22}.  
\par These detections, and non-detections, provide valuable insight for understanding the multivariate complexity of LyC escape. Various galaxy properties have been found to correlate (or anti-correlate) with the fraction of LyC photons escaping these galaxies ($\rm{f}_{\rm{esc,LyC}}$). Some of these include the observed UV-continuum slope ($\beta_{\rm{obs}}$, \cite{Chisholm22}), the star-formation rate surface density ($\Sigma_{\rm{SFR}}$, \cite{Flury22b}), the $[\text{O III}] \lambda 5007/ [\text{O II}] \lambda 3727$ (O32) ratio \citep{Nakajima14,Izotov18b} and strong C IV $\lambda \lambda$ 1548,1550 emission \citep{Saxena22}. However, given the complexity of the LyC escape from galaxies and the low-number statistics, these properties are at best thought to be necessary but insufficient conditions of LyC escape (e.g., \cite{Jung24}). Recently, \cite{Jaskot24}, using a multivariate analysis of detections and upper limits from the Low-redshift Lyman Continuum Survey (LzLCS, \cite{Flury22a}), attempted to rank these various properties based on their predictive power for LyC leakage. While the estimation of some of these properties poses a challenge at higher redshifts, properties like the observed UV-continuum slope can be estimated readily for galaxies well within the EOR \citep{Cullen23}.
\par Few LyC studies have probed the $z\sim 1-1.5$ redshift range due to a lack of UV sensitivity  and a historical difficulty in following up galaxies spectroscopically in the optical rest-frame. Combining 3D-HST grism spectroscopy (PI: P. van Dokkum, \cite{Momvheva16}) with GALEX UV imaging, \cite{Rutkowski16} were able to constrain the  $\rm{f}_{\rm{esc,LyC}}$ to $\lesssim 0.02$ for a sample of 600 $z\sim 1$ galaxies selected based on their H$\alpha$ emission. \cite{Alavi20} also derived constraints ($\rm{f}_{\rm{esc,LyC}}<0.06$) using the Solar Blind Channel on the HST with a sample of 11 SFGs at $1.2<z<1.4$ that were selected from the 3D-HST survey to have high H$\alpha$ equivalent widths ($>190 \text{\AA}$).

\par Recently, individual LyC leakers at $z\sim 1.1-1.5$ have been reported in the GOODS-South \citep{Saha20,Maulick24} and GOODS-North fields \citep{Dhiwar24} using the far-ultraviolet (FUV) F154W band of the UVIT. While filling a gap in the redshift space of LyC detections, these observations also represent some of the shortest wavelengths of LyC photons probed to date, reaching down to approximately $600 \text{\AA}$. The individual LyC detected galaxies are varied in terms of their morphology and properties. The galaxy detected in \cite{Saha20} is clumpy and is characterized by a high O32 ratio ($\sim  9.6$), large rest-frame [O III] and H$\alpha$ equivalent widths and a low-stellar mass ($M_{*}\sim 10^9 M_{\odot}$) whereas \cite{Maulick24} report the detection of LyC from a massive ($M_{*}\sim 10^{10.5} M_{\odot}$) dusty merging system with an extremely low O32 value (0.54). Both \cite{Saha20} and \cite{Dhiwar24} find evidence of LyC emission arising from regions that are offset from the optical center of the galaxies. 
\par In this work, we report the detection of 5 new LyC leaker candidates ($z\sim 0.99-1.42$) in the GOODS South Field. The LyC is detected in the F154W band image of the AstroSat UV Deep Field (AUDF) South survey \citep{Saha24}. We present our estimates of the galaxy properties that directly relate to ionizing photon escape or are considered indirect indicators, and we compare some of these with those of other LyC leakers reported in the literature. The paper is organized as follows: Section \ref{sec:data_sample} describes the procedure for selecting our sample; Section \ref{sec:phot} details the archival broadband imaging and catalogs, along with the methodology used for photometric measurements to assess the LyC detection significance of our sample. Section \ref{sec:properties} describes various properties of our sample from imaging and spectroscopic data, as well as those derived from spectral energy distribution (SED) modeling. Section \ref{sec:fesc_calc} outlines the estimation of quantities like the $\rm{f}_{\rm{esc,LyC}}$, that are directly related to the LyC escape. We discuss the implications of our findings in Section \ref{sec:discussion} and conclude in Section \ref{sec:summary}. All magnitudes quoted are in the AB system \citep{Oke83}. Throughout this work, we adopt the concordance $\Lambda$CDM cosmological model, with $H_0=70 \text{km}\text{s}^{-1}\text{Mpc}^{-1}$, $\Omega_m =0.3$, $\Omega_{\Lambda}=0.7$.
\section{Selection Strategy} \label{sec:data_sample}
We make a systematic search for LyC leakers as follows: we begin by querying sources in the GOODS South field that have the lines H$\beta$, [O III] and H$\alpha$ with S/N>3 in the publicly available HST grism catalogs released by the CLEAR survey \citep{Simons23}\footnote{\href{https://archive.stsci.edu/hlsp/clear}{https://archive.stsci.edu/hlsp/clear}}. The CLEAR survey targets the GOODS North and GOODS South fields using the HST WFC3 IR G102 grism and also combines existing spectroscopic data from other archival HST grism programs like that of the 3D-HST G141 grism observations \citep{Momvheva16}. The full datasets released by CLEAR thus include HST grism observations from 0.8-1.7 $\mu \rm{m}$.  
We constrain our objects to be in the redshift range $0.97< {z} < 1.55$. The lower redshift limit is chosen to ensure that any detected emission from these sources within the UVIT F154W band ($\sim 1340-1800 \text{\AA}$) corresponds to LyC photons ($<912 \text{\AA}$) only. The upper redshift limit is chosen to ensure that the H$\alpha$ line from galaxies in this redshift space will fall within the wavelength region of the grism G141 filter having a good throughput (beyond 1.65 $\mu$m the throughput falls rapidly). The selection of sources based on multiple emission lines ensures that the spectroscopic redshift is well determined. In addition to the CLEAR catalog, we query the recent JWST JADES \citep{Eisenstein23} NIRSpec Initial Data Release for the Hubble Ultra Deep Field DR3 redshift and emission-line flux catalog \citep{Bunker23}\footnote{\href{https://jades-survey.github.io/scientists/data.html}{https://jades-survey.github.io/scientists/data.html}} for sources that lie within our target redshift range. From this initial search we identify 137 galaxies. We then examine the positions of these sources in the AUDF South F154W band segmentation image \citep{Saha24} to look for objects detected by Source Extractor \citep{Bertin96} that lie within 1.6" ($\sim$ 1 PSF FWHM) from the candidates identified in the previous step.
This leaves us with 65 galaxies. We then carry out a visual inspection of the sources in the HST (F275W, F336W, F435W F606W, F814W) band images and reject sources that have one or more HST objects that lie within $\sim1.6$" from their center. This step is essential in any LyC search study using broad-band photometry given the susceptibility of foreground interlopers \citep{Siana07,Siana10,Vanzella10}. We are left with 7 candidate LyC leakers after this step. We then visually inspect the 1D spectra of the candidates (displayed in Appendix \ref{sec:1d_spec}) and discard one of the sources possessing broad-emission line features and that is characterized as hosting an active galactic nucleus (AGN) in the Chandra 7 Ms survey \citep{Luo17} based on its X-ray properties. Out of the remaining 6 candidates, one of them has been reported in \cite{Maulick24}. 
\par
\begin{figure*}[ht!]
\nolinenumbers
\includegraphics[width=1\textwidth]{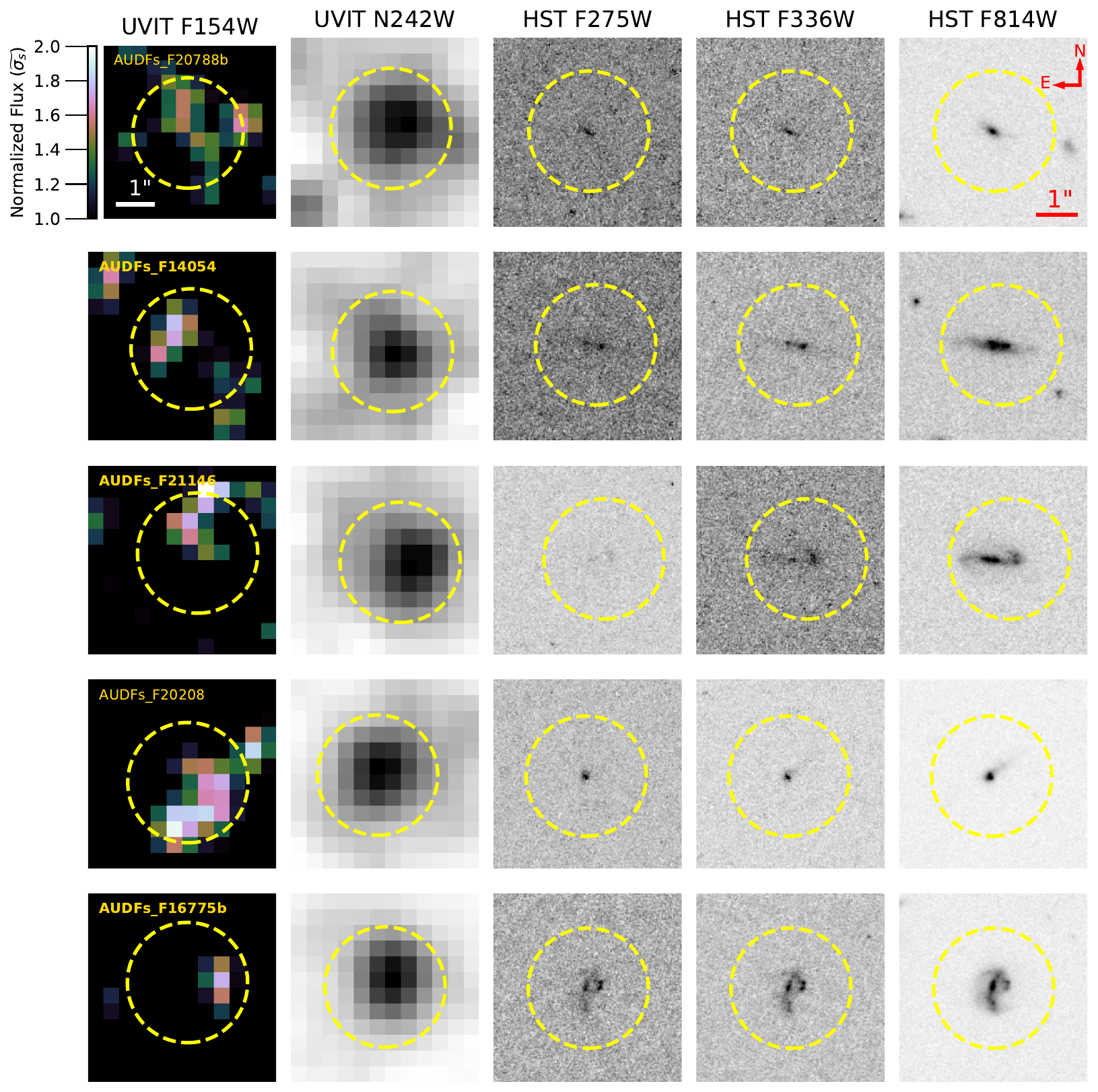}
\caption{Multiband imaging of our sample. Each galaxy corresponds to a row. The size of each postage stamp is $5^{''}\times 5^{''}$. The radius of the yellow dashed apertures is 1.6$^{''}$ ($\sim$ PSF FWHM of the F154W image) and they are centered on the HST F336W band centroids of the galaxies. The UVIT band images have been smoothed using a Gaussian kernel with a standard deviation of 1 pixel. The normalized fluxes in the F154W band images are indicated by the color-bar in the first row. The normalization is done with respect to the corresponding rms values in the regions of these objects. The  rms values, 1$\widetilde{\sigma_s}$ (from \cite{Saha24}) corresponding to the region in each of the five F154W images, from top to bottom are $[8.83,7.90,6.42,8.18,8.69]\times 10^{-6}\: \text{cnts/sec/pix}$. North is up and East is left in all the panels. The flux values in the UVIT N242W band are in units of cnts/sec/pix and in the HST band images they are in units of electrons/sec/pix.} 
\label{fig:stamp_images}
\end{figure*}

 Thus our final sample comprises of 5 new LyC leaker candidates. Figure \ref{fig:stamp_images} displays the 5 new identified LyC leaker candidates in the UVIT and HST broadband filters. The galaxy names \textbf{AUDFs\_F14054}, \textbf{AUDFs\_F21146}, and \textbf{AUDFs\_F16775b} are in bold throughout this work to distinguish them from other LyC leaker candidates in our sample, due to their unique morphology, which we discuss in Section \ref{sec:offset_lyc}.

\begin{figure}[ht!]
\nolinenumbers
\includegraphics[width=0.45\textwidth]{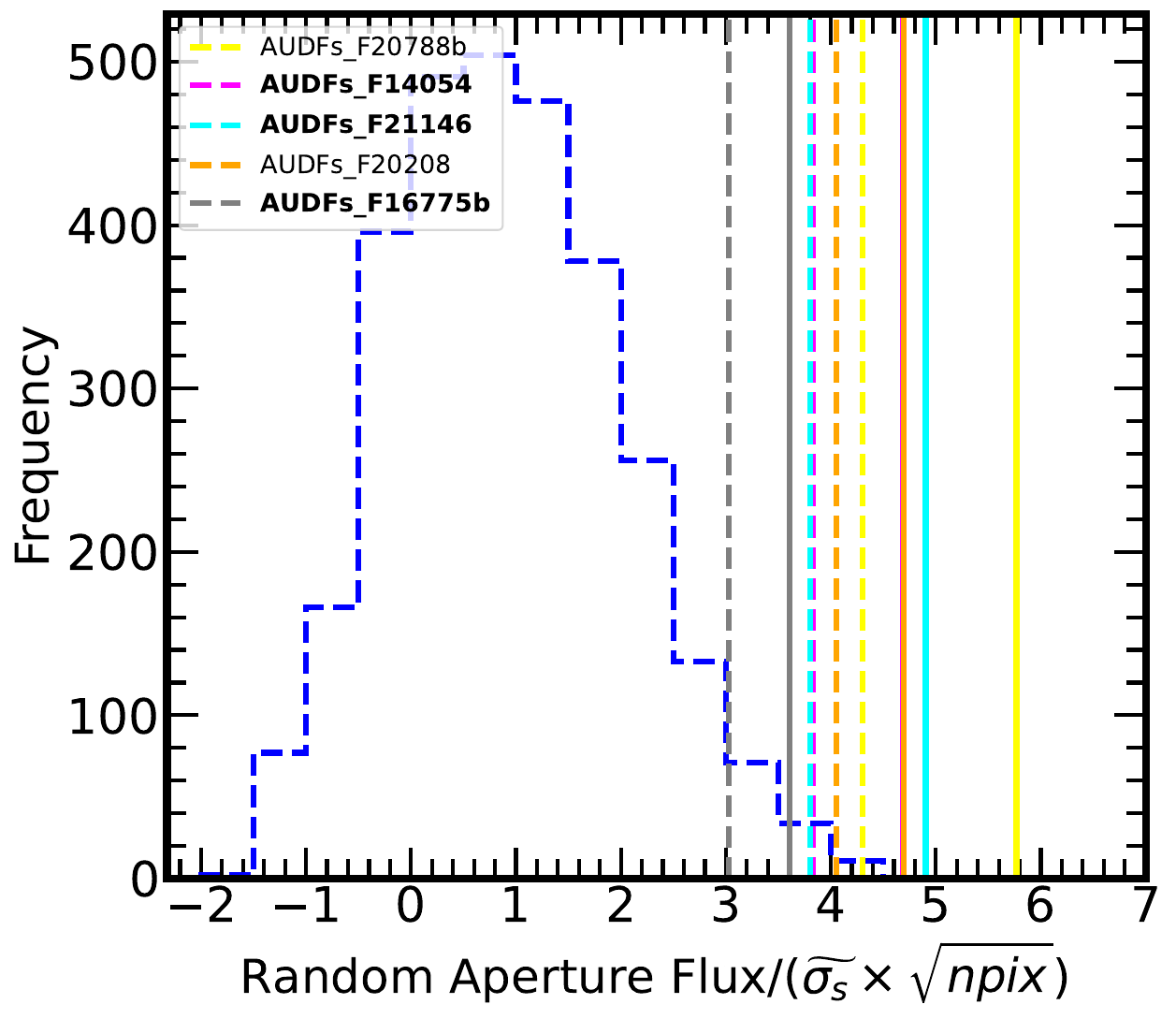}
\caption{The distribution of normalized fluxes (blue dashed histogram) within 3000 apertures of diameter PSF FWHM placed across empty regions of the AUDF South F154W image. The vertical colored dashed lines denote the normalized flux computed within the same sized aperture at the F336W centroids of the LyC leaker candidates identified in this work. {The solid colored lines denote the normalized flux when the apertures are placed at the F154W centroids (see Figure \ref{fig:seg_plot}).} The flux units are normalized to the background rms ($\widetilde{\sigma_s}$) and $npix$ denotes the number of pixels within the aperture.} 
\label{fig:empty_ap_flux}
\end{figure}

\begin{figure*}[ht!]
\nolinenumbers
\centering
\includegraphics[width=0.9\textwidth]{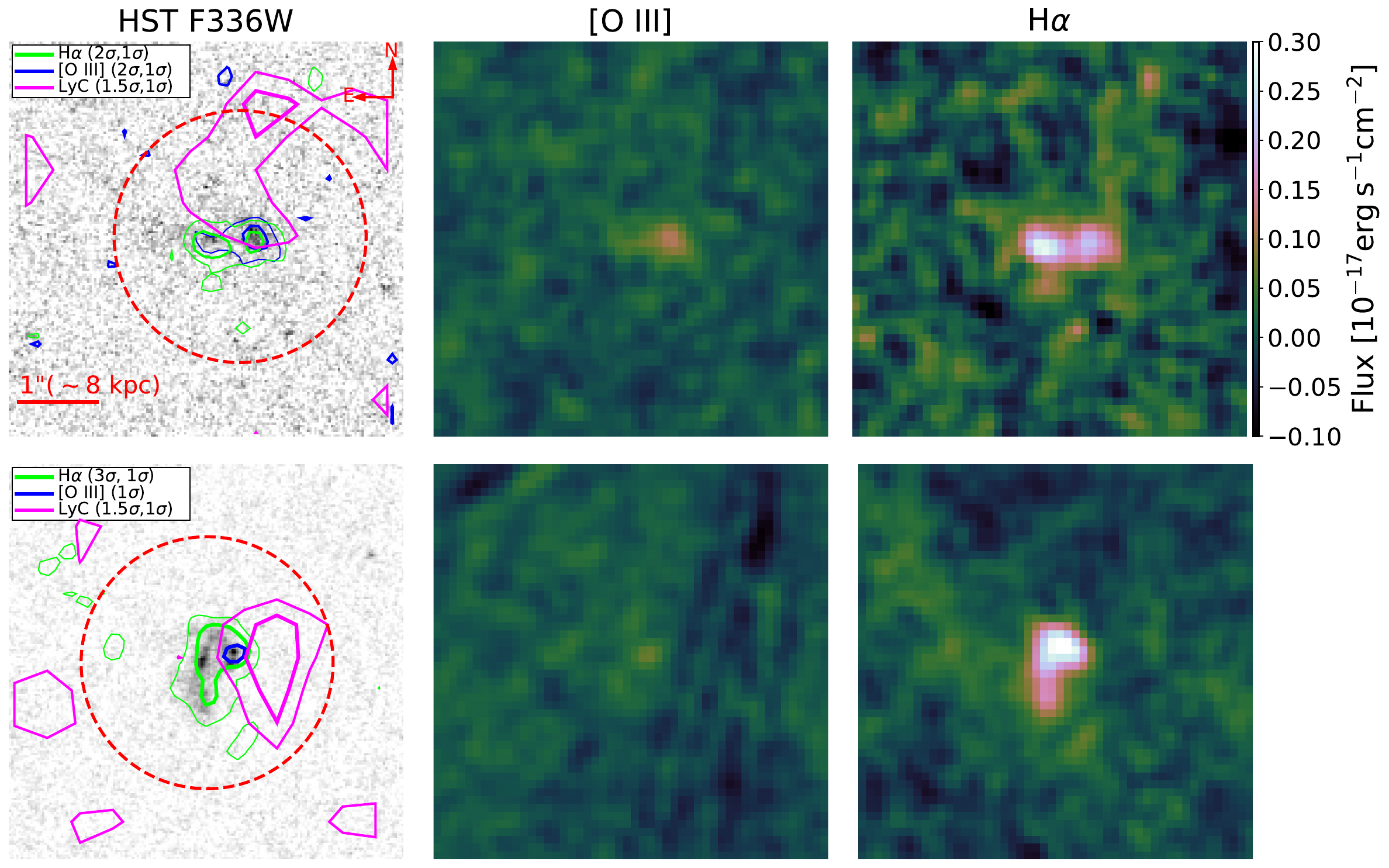}
\caption{From left to right: Column 1 displays the galaxies \textbf{AUDFs\_F21146} (top) and \textbf{AUDFs\_F16775b} (bottom) in the HST F336W band. Columns 2 and 3 display the [O III] and H$\alpha$ grism emission line maps respectively, of the galaxies obtained from the CLEAR survey \citep{Simons23}. The color bar in the first row corresponds to the fluxes for the [O III] and H$\alpha$ line maps. We overlay the H$\alpha$ (green), [O III] (blue) and LyC (magenta) emission contours on the F336W image in column 1. The thickness of the contour indicates its significance. The circular aperture is of radius 1.6$^{''}$ ($\sim$ PSF FWHM of the F154W band) and is centered on the F336W centroid of the objects. The pixel scale of the emission line maps is 0.1$^{''}$/pix. The emission line maps have been smoothed using a Gaussian kernel having a standard deviation of 1 pixel. The size of each image is $5^{''}\times 5^{''}$.} 
\label{fig:emi_maps}
\end{figure*}

\section{Imaging Data, detection significance and photometry} \label{sec:phot}
We use UVIT imaging \citep[AUDF South, see] []{Saha24} in conjunction with archival HST imaging from the {Hubble Legacy Field}\footnote{\href{https://archive.stsci.edu/prepds/hlf/}{https://archive.stsci.edu/prepds/hlf/}} \citep[HLF, see][]{Whitaker19} of the GOODS South Field to carry out the photometry of the selected galaxies that we subsequently use in the SED modeling. This imaging provides near continuous rest-wavelength coverage for $z\sim1-1.5$ galaxies between $\sim 750-7500 \text{\AA}$.
Before we proceed, we explain our methodology of evaluating the detection significance of the LyC signal in the UVIT F154W band.
\subsection{LyC Detection significance} \label{sec:LyC_sig}
Here, we perform forced aperture photometry on the sample in the UVIT F154W band image, using the centroids of the objects in the HST F336W band (HLF, \cite{Whitaker19}), to evaluate the significance of the LyC detection with respect to the variance in the image. The variance map is derived from the sigma map constructed in \cite{Saha24} (for a more detailed methodology about the background and variance map construction, see \cite{Pandey25}) and it reflects the differences in the exposure time in the different regions of the image. The typical rms value in the deepest regions of the AUDF South F154W image is $\sim 8.4\times10^{-6} \: \text{cnts/sec/pix}$, corresponding to a 3$\sigma$ depth of $\sim$ 27.9 magnitude when computed using a PSF FWHM diameter sized aperture. The aperture size we choose for this exercise is of diameter 1.6$^{''}$, which is the PSF FWHM of the UVIT F154W band. We normalize the fluxes with respect to the variance, $\widetilde{\sigma_s}^{2}$. The choice of the HST F336W centroid instead of the redder HST bands for these measurements is made since the F336W band roughly traces the rest-frame stellar FUV emission of these objects. Light from the older stellar populations traced by the redder bands would likely shift the centroids as is apparent for example in the galaxy \textbf{AUDFs\_F21146} in Figure \ref{fig:stamp_images}. We attempt to investigate the astrometric differences between the AUDF South F154W band and the HST HLF F336W band images.  Unfortunately, there is only a single star that is sufficiently bright in both the images. The difference in the centroid for this star between the two bands is 0.29$^{''}$  which is less than the pixel scale of the F154W band ($\sim 0.41$"/pix). 
\par We highlight the normalized F154W band fluxes (or significances) of the sample in Figure \ref{fig:empty_ap_flux}. . As in \cite{Saha24} and \cite{Maulick24} we also compare these
fluxes with those computed within apertures of the same size, placed randomly across empty regions of the AUDF South F154W image. Instead of 1000 apertures that were used in \cite{Saha24} and \cite{Maulick24}, we place 3000 (non-intersecting) apertures to better sample the AUDF South image. We display these fluxes in Figure \ref{fig:empty_ap_flux}. We note that 121 out of the 3000 apertures have a normalized flux or significance greater than 3. The distribution is skewed with respect to 0 and has a mean 0.73 and a standard deviation 1.10. We obtained similar deviations from the ideal expected mean of 0 and standard deviation of 1 in \cite{Saha24} (mean$\sim 0.8$, standard deviation $\sim 1.2$) for the 1000 apertures run (also see \cite{Rafelski15} for a discussion on this point). An 'empty' region in our experiment is defined in the context of the segmentation of sources that are detected using the Source Extractor parameters in \cite{Saha24}. Given this dependency on the Source Extractor parameters, we look for possible optical/IR counterparts that lie in the vicinity of these 'empty' positions using the HLF GOODS South photometric catalog \citep{Whitaker19}. 67 of the 121 'empty' positions where the flux significance is greater than 3, have an object in the HLF catalog within the PSF FWHM of the F154W band, i.e., 1.6$^{''}$. We acknowledge that a more complete analysis should take into account the brightness of these HST objects and their redshifts to ascertain their expected contribution in the observed F154W band (e.g., see \cite{Vanzella10}). However, as a cautionary measure, we provide the fraction of all empty apertures with significance exceeding that of the forced aperture photometric flux of the sample, which we denote by $\text{f}_{\text{EA}>\sigma_{\text{obj}}}$ in Table \ref{tab:cigale}, to emphasize that the calculated detection significance may not strictly conform to Gaussian statistical expectations. \par {We do note that the signal-to-noise ratio in the F154W band for our sample improves when the fluxes are measured within apertures centered on the actual F154W centroids of objects detected using Source Extractor (denoted by the cyan apertures in Figure \ref{fig:seg_plot}). Consequently, the signal-to-noise ratio derived from the forced aperture photometry using the F336W centroids represents a conservative estimate. For comparison, we mark the signal-to-noise derived from the F154W centroids by the solid colored lines in Figure \ref{fig:empty_ap_flux}.} 
\par For the SED modeling, we scale the F154W and N242W fluxes appropriately to account for the aperture correction \citep{Saha24}. 
\par {It is important to note that the false positives, by definition, are not detected by Source Extractor, whereas our identified candidates are. The Source Extractor parameters used in \cite{Saha24} require that a source must have at least 11 connected pixels with a threshold greater than 0.8$\sigma$ to be detected. For our candidates, we find that even with more conservative parameter choices—particularly when using a broader filter, which effectively suppresses noise peaks—these sources remain reliably detected (see Appendix \ref{sec:source_extractor} for details). The connectivity of the pixels and the detection of these sources after convolution with a broad filter further suggests that these are likely to be real sources. Along these lines, we propose an alternative definition of false positives in the context of Source Extractor detections in the F154W band with these parameters: F154W-detected sources for which there are no HST detected objects in the catalog of \cite{Whitaker19} within a PSF FWHM distance. 
Using this definition, we estimate the false positive density in the field to be 0.002 arcsec$^{-2}$ (see Appendix \ref{sec:source_extractor} for details).}

\subsection{Photometry from archival imaging}
Using the HST imaging obtained from the HLF \citep{Whitaker19}, we carry out aperture photometry within circular apertures of radius 1.4$^{''}$ placed at the coordinates (calibrated with respect to \textit{Gaia} DR2 \citep{GaiaDR2}) provided in the \cite{Whitaker19} catalog. The uncertainties in the fluxes are estimated using the corresponding weight images.
\par In addition to these two data sets, we adopt photometric measurements of these objects, carried out in the VLT/ISAAC Ks and Spitzer IRAC (Channels 1 to 4) images from the ASTRODEEP multi-wavelength photometric catalog of the GOODS South \citep{Merlin21}.

\begin{table*} 
\nolinenumbers
\hspace{-2 cm}\scalebox{0.73}{\begin{tabular}{cccccccccccccc}
\hline
\multicolumn{1}{c}{Object} & {\hspace{-0.75cm} \begin{tabular}[c]{@{}c@{}}RA\\ (J2000)\end{tabular}} & {\hspace{-0.75cm} \begin{tabular}[c]{@{}c@{}}Dec\\ (J2000)\end{tabular}} & \multicolumn{1}{c}{$z$}  & \multicolumn{1}{c}{\hspace{-0.75 cm}\begin{tabular}[c]{@{}c@{}}LyC detection\\significance\end{tabular}} & \multicolumn{1}{c}{\hspace{-0.75 cm}\begin{tabular}[c]{@{}c@{}}$\text{f}^{1}_{\text{EA}>\sigma_{\text{obj}}}$\end{tabular}} & \multicolumn{1}{c}{$\rm{M}_{\rm{UV}}$} & \multicolumn{1}{c}{$\beta_{\rm{obs}}$} &
\multicolumn{1}{c}{$(\rm{F}_{\rm{LyC}}/\rm{F}_{\rm{1500}})^{2}_{\rm{obs}}$} & \multicolumn{1}{c}{$\rm{f}^{\rm{H}\alpha \: \: 2}_{\rm{esc},\rm{LyC}}$} & \multicolumn{1}{c}{$\rm{f}^{\rm{SED}}_{\rm{esc},\rm{LyC}}$} & \multicolumn{1}{c}{$\text{log}_{10}(\xi_{\text{ion}}/\text{Hz\: erg}^{-1}$)} &
\multicolumn{1}{c}{$\text{log}(\rm{M}_{*}/\rm{M}_{\odot})$} & \multicolumn{1}{c}{\hspace{-0.75 cm} \begin{tabular}[c]{@{}c@{}}$\rm{SFR}_{\rm{SED}, 10\rm{Myr}}^{3}$\\ ($\rm{M}_{\odot} \text{yr}^{-1}$)\end{tabular}}  \\ \hline
AUDFs\_F20788b                     & 53.0559                                    & -27.7211                                        & 1.04                  & 4.30 & 0.002                                                                                          & -18.57                                       & $-1.86 \pm 0.10$ & $0.77 \pm 0.17$                                                                                             & $0.84 \pm 0.11$                                                                                                       & $0.66 \pm 0.14$       & 25.73                                                                                                           & $8.74 \pm 0.20$                                                                                  & $7.45 \pm 3.14$                                                                                                 \\ 
\textbf{AUDFs\_F14054}           & 53.1713                                    & -27.7930                                        & 0.99    &   3.83 & 0.008                                                                                           & -18.43                                       & $-1.36\pm  0.09$ & $0.52 \pm 0.14$                                                                                             & $0.59 \pm 0.17$                                                                                                       & $0.70 \pm 0.12$                                                                                                      & 25.16            & $9.23 \pm 0.02$                                                                                  & $4.59 \pm 0.20$                                                                                                 \\ 
\textbf{AUDFs\_F21146}                     & 53.1129                                    & -27.7199                                       & 1.11  & 3.80 & 0.009                                                                                          & -19.47                                        & $-1.06\pm  0.20$ & $0.25 \pm 0.06$                                                                                            & $0.14 \pm 0.09$                                                                                                      & $0.35 \pm 0.09$                                                                                                  & 25.14               & $9.98 \pm 0.03$                                                                                 & $24.60 \pm 1.22$                                                                                                \\ 
AUDFs\_F20208                     & 53.1281                                    & -27.7293                                        & 1.42 &  4.05 & 0.005                                                                                           & -20.14                                       & $-1.81\pm  0.06$ & $0.30 \pm 0.06$                                                                                             & $0.46 \pm 0.21$                                                                                                       & $0.50 \pm 0.06$   & 25.66                                                                                                               & $9.31 \pm 0.11$                                                                                  & $30.07 \pm 3.08$                                                                                                 \\ 
\textbf{AUDFs\_F16775b}                     & 53.1588                                    & -27.7706                                        & 1.00  &  3.03 & 0.034                                                                                           & -19.13                                       & $-0.88\pm  0.17$ & $0.34 \pm 0.10$                                                                                             & $0.27 \pm 0.10$                                                                                                       & $0.32 \pm 0.12$   & 25.12                                                                                                              & $9.85 \pm 0.02$                                                                                  & $16.94 \pm 0.86$    \\   

\end{tabular}}
\caption{Relevant information regarding the selected sample and properties derived from the SED fitting.\\ 
Notes: \\
\small{$^1$ Fraction of 'empty' apertures with flux significance greater than the LyC detection significance of the object (see text for details). \\
$^2$ Corrected for IGM attenuation (See Section \ref{sec:fesc_Halpha}). \\
$^3$ SFR averaged over the last 10 Myr.}} \label{tab:cigale} 
\end{table*}

\begin{table*} 
\nolinenumbers

\hspace{-2cm}\scalebox{0.74}{\begin{tabular}{ccccccccccccc}
\hline
\multicolumn{1}{c}{Object} & \multicolumn{1}{c}{CLEAR ID} &  \multicolumn{1}{c}{MUSE ID$^{1}$} &  \multicolumn{1}{c}{$\rm{F}_{[\rm{O II} \lambda 3727]}$} & \multicolumn{1}{c}{$\rm{F}_{\rm{H}\beta}$} & \multicolumn{1}{c}{$\rm{F}_{[\rm{O III}]}^{2}$} & \multicolumn{1}{c}{$\rm{F}_{\rm{H}\alpha}$} & \multicolumn{1}{c}{$\text{EW}_{\text{H}\beta,\text{rest}} (\text{\AA})$} &
\multicolumn{1}{c}{$\text{EW}_{\text{[OIII]},\text{rest}} (\text{\AA})$} &   
\multicolumn{1}{c}{$\text{EW}_{\text{H}\alpha,\text{rest}} (\text{\AA})$} & \multicolumn{1}{c}{$\rm{E(B-V)}_{\rm{neb}}$} & \multicolumn{1}{c}{\hspace{-0.75 cm}\begin{tabular}[c]{@{}c@{}}$\rm{SFR}_{\rm{H}\alpha}^{3}$\\ ($\rm{M}_{\odot} \text{yr}{-1}$)\end{tabular}} &  \multicolumn{1}{c}{Class$^{4}$}  \\ \hline
AUDFs\_F20788b                      & 40553                                    & -                                                   & -                                                                                           & $2.05 \pm 0.42$                                       & $8.29 \pm 0.48$ & $6.15 \pm 0.45$                                                                                             & $66.22^{+\tiny{13.57}}_{-\tiny{13.56}}$                                                                                                       & $214.67^{+\tiny{14.12}}_{-\tiny{13.66}}$                                                                                                                  & $308.85^{+\tiny{22.41}}_{-\tiny{24.04}}$                                                                                  & $0.02 \pm 0.26$  & $2.97 \pm  1.89$ & SF                                                                                                \\ 
\textbf{AUDFs\_F14054}                      & 25304                                    & 1002                                                        & $1.38 \pm 0.01$                                                                                           & $0.18 \pm 0.04$                                       & $1.13 \pm 0.05$ & $1.26 \pm 0.03$ & -                                                                                                       & -                                                                                                                  & -                                                                                 & $1.02 \pm 0.26$ & $6.37 \pm 4.09$ & AGN                                                                                                 \\ 
\textbf{AUDFs\_F21146}                     & 40878                                    & -                                                     & -                                                                                          & $2.22 \pm 0.56$   & $6.14\pm  0.64$ & $15.28 \pm 1.07$                                                                                            & $16.15^{+\tiny{3.96}}_{-\tiny{3.94}}$                                                                                                                 & $37.09^{+\tiny{3.85}}_{-\tiny{4.10}}$ & $140.02^{+\tiny{10.38}}_{-\tiny{10.26}}$                                                                                 & $0.93 \pm 0.30$  & $75.96 \pm 56.29$ & SF                                                                                               \\ 
AUDFs\_F20208                     & 38849                                    & -                                                        & -                                                                                           & $3.43 \pm 1.01$                                       & $24.30\pm  1.06$ & $2.97 \pm 0.66$  & $61.10^{+\tiny{18.04}}_{-\tiny{17.93}}$                                                                                                                  & $370.49^{+\tiny{20.27}}_{-\tiny{19.74}}$                                                                                  & $395.07^{+\tiny{22.30}}_{-\tiny{23.16}}$ & $0.32 \pm 0.34$ & $28.53 \pm 23.87$ & AGN                                                                                                \\ 
\textbf{AUDFs\_F16775b}                     & 30520                                    & 13                                                      & $3.71 \pm 0.02$                                                                                           & $4.22 \pm 0.55$   & $4.69 \pm 0.58$ & $20.94 \pm 0.65$                                                                                             & $20.67^{+\tiny{2.84}}_{-\tiny{2.71}}$                                                                                                                 & $19.07^{+\tiny{2.30}}_{-\tiny{2.31}}$                                                                                  & $139.06^{+\tiny{4.50}}_{-\tiny{4.61}}$ & $0.54 \pm 0.15$ & $30.30 \pm 11.25$ & SF   \\
\end{tabular}}
\caption{Spectroscopic information of the sample from the literature \citep{Simons23,Bacon23,Bunker23} and its related derived properties. The line fluxes in Columns 4-7 are in units of  $10^{-17}\text{erg}\ \text{s}^{-1}\ \text{cm}^{-2}$ and represent the observed values. Thus they have not been corrected for internal extinction. The rest-frame optical line fluxes of \textbf{AUDFs\_F14054} are adopted from the JADES DR3 NIRSpec medium gratings catalog of \cite{Bunker23}, whereas for the rest of the galaxies these are from the CLEAR HST grism catalog \citep{Simons23}. Whenever available, the [O II $\lambda$3727] line flux is from the MUSE Hubble Deep Field Surveys DR2 catalog \citep{Bacon23}.\\ 
Notes: \\
\small{$^1$ ID in the MUSE HUDF DR2 catalog \citep{Bacon23}. \\
$^2$ $\rm{F}_{[\rm{O III}]}$ denotes the flux of the entire [O III] doublet since it is unresolved in the HST grism data. For \textbf{AUDFs\_F14054} however, this value corresponds to the $\lambda 5007$ line of the doublet since the medium resolution NIRSpec grating is capable of resolving the doublet. \\$^3$ The $\rm{SFR}_{\rm{H}\alpha}$ values have been corrected for internal extinction. \\ $^4$ The location of these objects in the BPT diagram adopted from \cite{Kewley01} (see Figure \ref{fig:bpt_sii}).}} \label{tab:spec_prop}
\label{tab:info}
\end{table*}

\begin{table}
\nolinenumbers
\hspace{-1cm}\scalebox{0.7}{\begin{tabular}{c|c|c} 

\hline
Module                                                                               & Parameter                                                                                              & Parameter input                                                                                     \\ \hline
{sfh2exp}                                                             & $\tau_{\text{main}}^1$ (Myr)                                                                             & \begin{tabular}[c]{@{}c@{}}100,500,1000,1500,\\ 3000,4000\end{tabular}  \\   & $\tau_{\text{burst}}^2$ (Myr)                                                                            & 50,100,500,1000,2500   \\ & $\text{f}_{\text{burst}}^3$  & 0.1,0.2,0.3,0.4,0.5,0.6 \\ & $\text{age}_{\text{main}}^4$ (Myr)  & \begin{tabular}[c]{@{}c@{}}200,300,500,1000,\\ 1500,2000,3000,5000\end{tabular}  \\ & $\text{age}_{\text{burst}}^5$ (Myr) & 2,5,10,20,40,50  \\ \hline &    &   \\ 
{bc03}  & IMF  & {Salpeter (0)}  \\  & metallicity & \begin{tabular}[c]{@{}c@{}}0.0001,0.0004,0.004,\\ 0.008,0.02\end{tabular} \\ \hline & &  \\ {nebular} & {Gas metallicity} & \begin{tabular}[c]{@{}c@{}}0.0001,0.0004,0.001,0.004, \\0.008,0.02\end{tabular} \\ & {$f_{\rm{esc},\text{LyC}}$} & \begin{tabular}[c]{@{}c@{}}0.05,0.1,0.2,0.3,0.4,0.6,\\ 0.7,0.8\end{tabular} \\ & \begin{tabular}[c]{@{}c@{}}fraction of LyC photons\\ absorbed by dust\\ $f_{\text{dust}}$\end{tabular} & 0,0.1,0.2  \\ \hline &  &  \\ 
{Calzetti dust attenuation} & $\text{E(B-V)}_{\text{gas}}$                                                                                  & \begin{tabular}[c]{@{}c@{}}0.05,0.15,0.25,0.40,\\ 0.50,0.70,0.90\end{tabular} \\  & $\text{E(B-V)}$ factor$^{7}$    & 0.44    \\  & $\delta^{6}$ & -0.5 \\ &  {UV bump wavelength} & 217.5 nm   \\  &    &  \\ \hline {restframe parameters}   & {UV $\beta$ slope (Calzetti)}  & True  \\   & Dn4000$^{8}$   & True \\  & {Infrared excess (IRX)} & True  \\ 
\end{tabular}}
\caption{Some of the modules and associated parameters used in the CIGALE modeling.\\
Notes:
\small{\\$^1$ The e-folding time of the main stellar population model.
\\$^2$ The e-folding time of the late starburst population.
\\$^3$ Mass fraction of the starburst population.
\\$^4$ Age of the main stellar population in the galaxy.
\\$^5$ Age of the late starburst.
\\$^6$ Modification of the attenuation curve.
\\$^7$ Factor used to convert $\text{E(B-V)}_{\text{gas}}$ to $\text{E(B-V)}_{\text{stellar}}$.
\\$^8$ Application of the 4000 $\text{\AA}$ break \citep{Balogh99}.}} \label{tab:cigalegrid}
\end{table}

\begin{figure*}[ht!]
\nolinenumbers
\centering
\includegraphics[width=2\columnwidth]{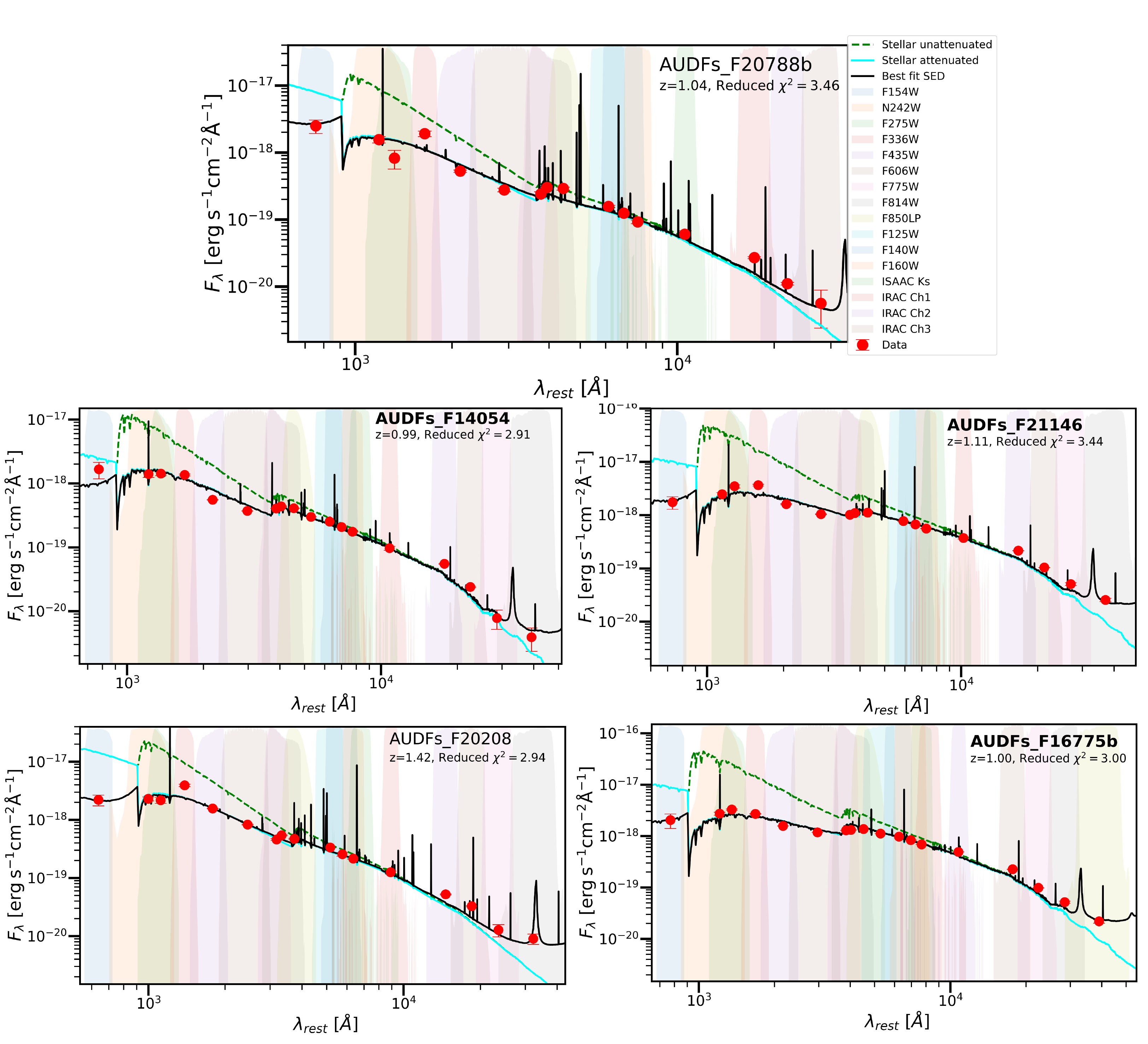}
\caption{The best fit CIGALE SED models (black curve) and photometric data (red points) of the five new identified LyC leaker candidates in this work. The green dashed and cyan curves denote the stellar unattenuated and stellar attenuated best fit CIGALE models respectively. The colored curves in the background represent the shapes of the photometric filters used. The reduced $\chi^2$ values of the best-fits SEDs are indicated in the text in the top-right.} 
\label{fig:galaxy1_sed}
\end{figure*}

\section{Properties of the sample} \label{sec:properties}
The combination of imaging and spectroscopy allows us to characterize the stellar and nebular properties of these candidate LyC leakers. We first use the available high-resolution HST imaging to highlight the morphology of some of the more extended galaxies in our sample in which we find tentative evidence of the LyC emission having significant spatial offsets from the non-ionizing UV continuum centers. We then outline our spectral energy distribution (SED) modeling and the properties derived from it.

\subsection{Clumpy morphology and spatially offset LyC emission} \label{sec:offset_lyc}
The multiband HST imaging in Figure \ref{fig:stamp_images} reveals clumpy regions in the extended galaxies \textbf{AUDFs\_F14054}, \textbf{AUDFs\_F21146} and \textbf{AUDFs\_F16775b} that are relatively bluer than the rest of the galaxy. 
The F275W and F336W bands probe the rest-frame FUV emission of these galaxies at their redshifts. The blue clumpy features are then possible tracers of the relatively less-obscured/younger star-forming regions in these galaxies. In the redder bands, \textbf{AUDFs\_F21146} has the morphology of a spiral galaxy (F814W band in Figure \ref{fig:stamp_images}), with the light being dominated by the older stellar populations. 
\par A natural concern that arises is whether the clumps in these objects could be foreground interlopers. To this end, we try to estimate the photometric redshifts of the clumps by performing PSF photometry within apertures of radius 0.2" centered on the clumps in the HST band images and, we examine the HST grism emission line maps of these galaxies provided in the data release of the CLEAR survey \citep{Simons23}. We further note that the visual inspection of these galaxies in multiple HST bands (Figure \ref{fig:stamp_images}) reveals signatures of interaction between the clumps and the main body of the galaxies. 
\par We use the code \texttt{EAZYpy} \citep{Brammer08,Brammer21} for the estimation of the photometric redshifts of the clumps and include a brief description of our methodology in Appendix \ref{sec:eazy}. The photometric redshifts we obtain for the clumps belonging to \textbf{AUDFs\_F14054} (CF14054) and \textbf{AUDFs\_F21146} (CF21146) are in excellent agreement with the spectroscopic redshifts. For the clumps associated with \textbf{AUDFs\_F16775b} (CF16775b) we obtain a photometric redshift of $0.72^{+0.36}_{-0.26}$ while the spectroscopic redshift is 1.00. If the redshift of C16775b is indeed $<0.97$, it is important to note that the flux in the F154W band would be contaminated by non-ionizing radiation. However, given the morphology in the H$\alpha$ map (discussed below), the visual inspection of the object in the other HST bands, and the lack of unidentified emission lines in the spectra, we find it unlikely that this is a foreground interloper. We present additional spectral data for this galaxy in Appendix \ref{sec:galaxy5_spec}, which supports our hypothesis of the absence of a foreground interloper in the system.
\par The emission line maps were constructed by \cite{Simons23} using the grism redshift and line analysis software \texttt{grizli}\footnote{\href{https://github.com/gbrammer/grizli/}{https://github.com/gbrammer/grizli/}} \citep{Brammer19}. The continuum and contamination subtracted 2D spectral beams were drizzled using the astrometry of the slitless exposures projected along the spectral trace to the redshifted line center wavelength. The emission line maps have an effective spatial resolution of 0.16" \citep{Simons21}. For more details we refer the reader to \cite{Simons21,Simons23}. The emission line maps of \textbf{AUDFs\_F14054} have low S/N, and its associated clump remains unresolved. Therefore, we exclude \textbf{AUDFs\_F14054} from the following discussion on the emission line maps.
The H$\alpha$ (blended with [N II] in the grism) emission line maps (third column, Figure \ref{fig:emi_maps}) of \textbf{AUDFs\_F21146} and \textbf{AUDFs\_F16775b} reveal line emission from both the main body and the clumpy regions of these galaxies, thus providing strong evidence that these clumps are at the same redshift as that of the galaxy. In the case of \textbf{AUDFs\_F21146}, only the western blue clump seems to be emitting [O III]. This suggests that this region hosts harder ionizing sources than the rest of the galaxy. 
\par The LyC emission traced by the UVIT F154W band shows signatures of being offset from both the stellar FUV continuum and the line emission in the galaxies \textbf{AUDFs\_F21146} and \textbf{AUDFs\_F16775b} (Column 1, Figure \ref{fig:emi_maps}). In physical scales these offsets correspond to distances of $\sim 9$ and $\sim 5$ kpc for the galaxies  \textbf{AUDFs\_F21146} and \textbf{AUDFs\_F16775b} respectively. Considering an astrometric uncertainty of 1 pixel in the F154W band ($\sim$0.4"), the uncertainties in the physical scale correspond to $\sim 3.4$ kpc.
However, it is also important to note that this inference is based on the LyC emission that is largely confined within the PSF FWHM of the F154W band (1.6"), as indicated by the red aperture in Figure \ref{fig:emi_maps}.  We return to spatially offset LyC emission in Section \ref{sec:discussion}.

\subsection{Spectral Energy Distribution fitting} \label{sec:sed}
We employ the code CIGALE \citep{Boquien19, Yang22} for the SED-fitting. Our full grid of models is defined in Table \ref{tab:cigalegrid}.
Specifically, we adopt the the Bruzual and Charlot \citep{Bruzual03} single stellar population library (\textit{bc03}) with a Salpeter IMF \citep{Salpeter55} and model the SFH using CIGALE's \textit{sfh2exp} module, which accounts for the primary stellar population formed at an age $\gtrsim 200\: \rm{Myr}$ and a secondary, more recent burst with age $\lesssim 50\: \rm{Myr}$.
For modeling the dust attenuation we adopt a modified version of the Calzetti starburst attenuation law \citep{Calzetti00}. We use the power-law modification index $\delta=-0.5$ which results in a curve that resembles the SMC extinction curve.
For more details about the parameters we refer the reader to \citet{Boquien19}. We include some of the best fit parameters and physical estimates from the SED modeling in Table \ref{tab:cigalegrid} and we display the best-fit SED plots in Figure \ref{fig:galaxy1_sed}. We note that CIGALE does not attenuate the stellar light at wavelengths less than 912 $\text{\AA}$. This explains why the intrinsic and dust-attenuated stellar SEDs are the same for the ionizing part of the spectrum in Figure \ref{fig:galaxy1_sed}. 

\subsection{Stellar mass, $M_{\text{UV}}$, $\beta$ and $\xi_{\rm{ion}}$}
The best-fit CIGALE model stellar masses of our sample span about 1 dex, from $\sim 10^9-10^{10}\: \rm{M}_{\odot}$. We estimate the absolute observed UV magnitudes ($\rm{M}_{\rm{UV}}$) at 1500 $\text{\AA}$ and the observed UV-continuum slope ($\beta_{\rm{obs}}$) for our sample from the best fit CIGALE SEDs. $\beta_{\rm{obs}}$ is estimated in the range 1268–2580 \text{\AA} \citep{Calzetti00}. The majority of our sample comprises of galaxies $\gtrsim\text{M}^{*}_{1500}\sim -18.5$, estimated in the literature for galaxies (excluding AGN) at $z \sim 0.7-1$ \citep{Bhattacharya23,Sun23}. We compare our estimated values for these quantities with those from other LyC leaker studies in Section \ref{sec:discussion}. We also estimate the ionizing photon production efficiency ($\xi_{\rm{ion}}$) for our sample using the intrinsic non-attenuated stellar SED output from the best-fit CIGALE SEDs. This quantity is defined as the following, $\xi_{\rm{ion}}=\frac{\dot{N}^{int}_{\rm{LyC}}}{L_{1500}}$, where $\dot{N}^{int}_{\rm{LyC}}$ denotes the intrinsic ionizing photon production rate and $L_{1500}$ denotes the intrinsic luminosity density at rest-frame $1500\: \text{\AA}$ in units of $\text{erg}\  \text{s}^{-1} \text{Hz}^{-1}$ for a galaxy. The $\text{log}_{10}(\xi/\text{Hz\: erg}^{-1})$ values for our sample span the range $25.12-25.73$, thus lying either around or exceeding the canonical values typically used in reionization scenarios  ($\text{log}(\xi_{\rm{ion}}/\text{Hz\: erg}^{-1})\simeq 25.2-25.3$, \cite{Robertson13}). 
\subsection{Dust attenuation} \label{sec:dust}
We estimate the nebular dust attenuation of the galaxies in our sample using the Balmer decrement. The H$\beta$ and H$\alpha$ line fluxes are adopted from the CLEAR catalog \citep{Simons23}. Since the H$\alpha$ and [N II] doublet are merged in the grism G141 spectra, we employ an empirical correction for the [N II] contamination using the observed equivalent width of the merged line following \cite{Sobral12}. Assuming a temperature $T=10^4$K, we adopt an intrinsic Balmer decrement of 2.87. We subsequently estimate the nebular reddening $\rm{E(B-V)_{neb}}$ and the dust-corrected luminosity of the H$\alpha$ line. To this end, we adopt the nebular extinction curve of \cite{Reddy20}.
\par We estimate the star-formation rates ($\text{SFR}_{\text{H}\alpha}$) for our sample using the dust-corrected luminosity of the H$\alpha$ lines adopting the \cite{Kennicutt98} calibration for a Salpeter IMF. The $\text{SFR}_{\text{H}\alpha}$ is a tracer of the recent star-formation in the galaxy, since the Balmer emission line fluxes are dominated by young massive stars having masses$>10 M_{\odot}$ and lifetimes $<20$ Myr \citep{Kennicutt98}. Within the 1$\sigma$ error, albeit the errors are large, our $\text{SFR}_{\text{H}\alpha}$ values are in agreement with the $\rm{SFR}_{\rm{SED}, 10\rm{Myr}}$ values obtained from the best fit SED models. 

\subsection{AGN diagnostics}
\begin{figure}[ht!]
\nolinenumbers
\includegraphics[width=0.5\textwidth]{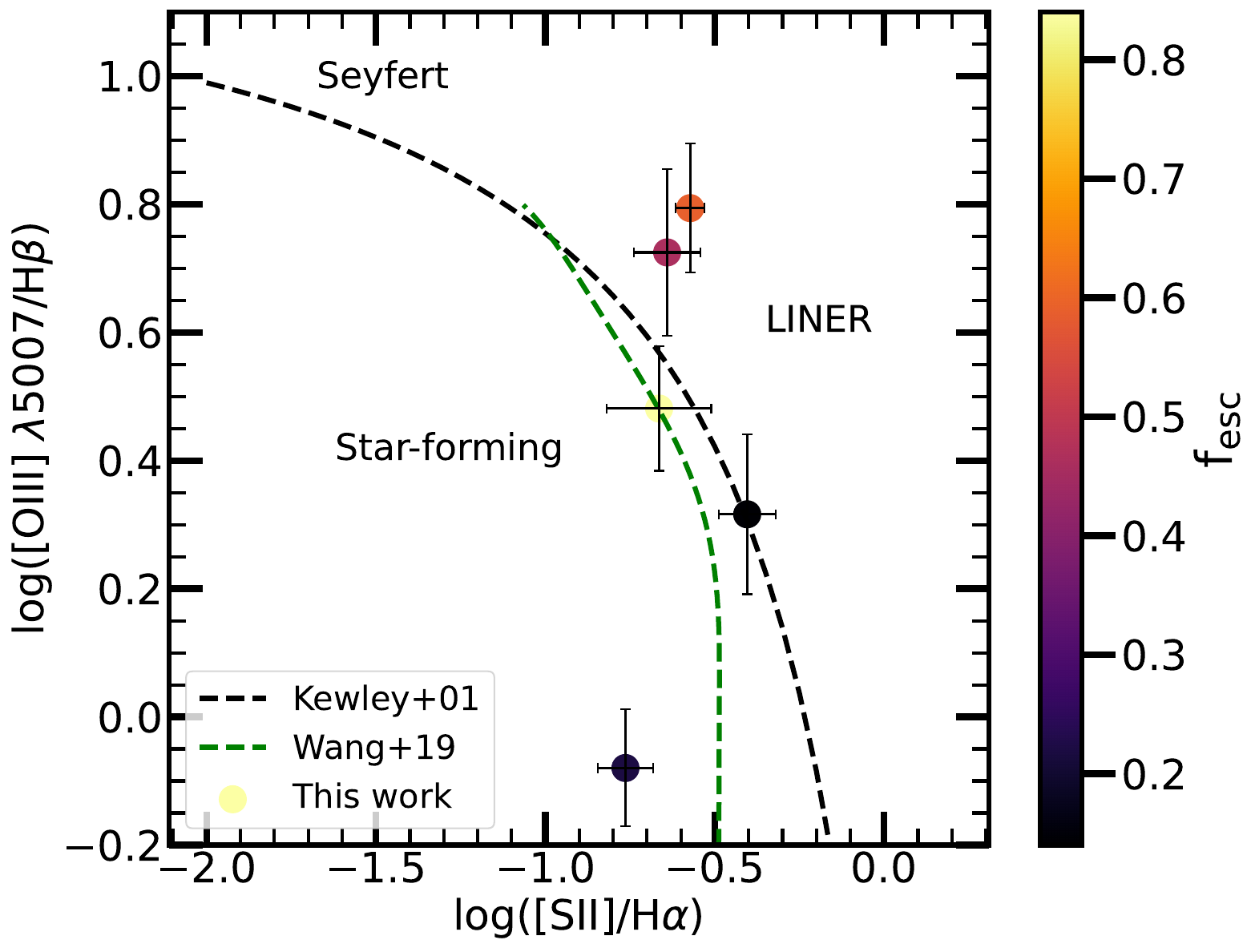}
\caption{The location of our sample in the BPT diagnostic plot adapted from \cite{Kewley01}. The green dashed curve indicates the plot used by \cite{Wang19} to quantify the [S II] deficiency.} 
\label{fig:bpt_sii}
\end{figure}

None of the galaxies in our sample have been detected in the X-ray in the Chandra 7 Ms survey \citep{Luo17}. We utilize the line flux measurements from the CLEAR \cite{Simons23} and JADES NIRSpec \citep{Bunker23} catalogs to locate our sample on the $\rm{log}([\rm{S\:II}]/\rm{H}\alpha)$ vs $\rm{log}([\rm{O\:III}]/\rm{H}\beta)$ Baldwin, Philips, \& Terlevich (BPT, \cite{Baldwin81}) diagnostic diagram adapted from \cite{Kewley01} (Figure \ref{fig:bpt_sii}). Since the [O III] $\lambda \lambda$ 4959,5007 doublet is unresolved in the grism spectra, we adopt the theoretical line ratio of 2.98 for the [O III] doublet \citep{Storey00} to separately estimate the flux of the [O III] $\lambda$5007 line. \textbf{AUDFs\_F14054} and AUDFs\_F20208 both lie in the region between Seyferts and LINERS whereas AUDFs\_F20788b and \textbf{AUDFs\_F16775b} lie in the star-forming region of the plot. \textbf{AUDFs\_F21146} lies nearly on the extreme starburst line of \cite{Kewley01}.

\section{Escape of ionizing photons} \label{sec:fesc_calc}
In this section we quantify the escape of ionizing photons from our sample. It is worth noting that the LyC wavelengths probed in this work ($\sim 630-775 \text{\AA}$) are shorter than the typical LyC wavelengths probed in other studies ($\sim 900 \text{\AA}$, e.g., \cite{deBarros16,Izotov18,Flury22a}). 
\par We first compute the ratio of the ionizing flux density probed by the F154W band to the observed non ionizing flux density at 1500 \text{\AA} \big($(\rm{F}_{\rm{LyC}}/\rm{F}_{\rm{UV}})_{\rm{obs}}$\big). This quantity is largely model independent and hence can be compared readily across surveys. We next outline our methodology for estimating the LyC escape fraction $f_{\rm{esc},\text{LyC}}$. In the literature, there are various approaches to estimating the $f_{\rm{esc},\text{LyC}}$, each with slightly different definitions and assumptions. For instance, estimating the so called relative escape fraction entails quantifying the escape of LyC photons at a monochromatic wavelength, usually at $\sim 900$ \text{\AA} relative to the production and escape of photons at $1500$ \text{\AA} \citep{Steidel18}. The second is that of the absolute escape fraction which is a simple measure of the fraction of all LyC photons escaping into the IGM without being absorbed by the HI and dust within the galaxy. In the following, our estimates of the escape fraction are closer to the definition of the absolute escape fraction though bearing in mind certain caveats. \\
\subsection{LyC to non-ionizing UV flux ratio}
As was done for the estimation of $\rm{M}_{\rm{UV}}$, we estimate the non-ionizing flux density at 1500 \text{\AA} from the best-fit CIGALE SED. The observed LyC to non-ionizing flux density ratio is readily comparable to other LyC studies due to its widespread adoption in the literature \citep{Steidel18,Wang19,Flury22a,Kerutt23,Dhiwar24} and its direct nature. It does not require any assumptions about the dust, albeit the IGM attenuation does need to be taken into account when comparing leakers at different redshifts.  We discuss this further in Section \ref{sec:discussion}.

\subsection{$f_{\rm{esc},\text{LyC}}$ from the dust-corrected H$\alpha$ luminosity} \label{sec:fesc_Halpha}
Our approach to estimating the Lyman-continuum escape fraction ($f_{\rm{esc},\text{LyC}}$) in this sub-section follows that of \cite{Saha20,Dhiwar24} and \cite{Maulick24}. We first convert the dust-corrected luminosity 
of the H$\alpha$ line into the non-escaping ionizing photon rate  $\dot{N}^{non-esc}_{\rm{LyC}}$. Assuming a temperature $T=10^4$K and electron density $n_e=100\: \text{cm}^{-3}$, this is given by \citep{Ferland06},
\begin{equation}  \label{eq:N_nonesc}
    \dot{N}^{non-esc}_{\rm{LyC}} [\:\text{s}^{-1}]=7.28 \times 10^{11} L_{int,\rm{H}\alpha} [\text{erg}\: \text{s}^{-1}] .
\end{equation}
where $ L_{int,\rm{H}\alpha}$ is the dust-corrected H$\alpha$ luminosity estimated in Section \ref{sec:dust}. We estimate the rate at which LyC photons escape the galaxy, $\dot{N}^{esc}_{\rm{LyC}}$ from the observed F154W flux. Following \cite{Maulick24}, this is given by,     \begin{equation}\label{eq:NescLyC} 
    \dot{N}^{esc}_{\rm{LyC}} = \frac{L_{\rm{F154W}} \times \lambda^{\rm{F154W}}_{\rm{rest}}}{hc/\lambda^{\rm{F154W}}_{\rm{rest}}} \times \rm{exp}[{\tau_{\rm{IGM}}}] ,
\end{equation}
where $\lambda^{\rm{F154W}}_{\rm{rest}} = 1541\text{\AA}/(1+z)$ for the F154W filter and $L_{\rm{F154W}} \times \lambda^{\rm{F154W}}_{\rm{rest}}$ denotes the rest-frame luminosity in units of $\text{erg}\:\text{s}^{-1}$ corresponding to the photons captured in the F154W band. 
Given $\dot{N}^{non-esc}_{\rm{LyC}}$ and $\dot{N}^{esc}_{\rm{LyC}}$, we then estimate the escape fraction using the following equation, 
\begin{equation} 
  f^{\rm{H}\alpha}_{\rm{esc,LyC}}=\frac{\dot{N}^{esc}_{\rm{LyC}}}{\dot{N}^{esc}_{\rm{LyC}}+\dot{N}^{non-esc}_{\rm{LyC}}}.
  \label{eq:fesc}
\end{equation}
Note that the attenuation of the LyC flux by the IGM is implicitly included in the term, $\dot{N}^{esc}_{\rm{LyC}}$. We adopt a mean IGM transmission of 0.63 in the F154W band, based on Monte Carlo simulations by \cite{Dhiwar24} using the column-density distribution from \cite{Inoue14}. These simulations, performed for four objects at redshift $z\sim1.2$, included 10000 IGM sightlines per object. 
We note that our definition of $f^{\rm{H}\alpha}_{\rm{esc,LyC}}$ is inherently a lower limit to the absolute escape fraction since it takes only those escaping ionzing photons into account that are probed by the F154W band.
 \subsection{$f_{\rm{esc},\text{LyC}}$ from the best fit SED}
In the SED fitting, we allow for a range of $f_{\rm{esc},\text{LyC}}$ values in the CIGALE module \textit{nebular}. This module computes the nebular line and continuum emission based on templates generated using CLOUDY 13.01 \citep{Ferland98,Ferland13}. The module also incorporates the absorption of a fraction of the intrinsic ionizing photons by the dust ($f_{\rm{dust}}$). This has an effect in the strength of the nebular emission lines in the model observed SED. It is worth reiterating that the intrinsic ionizing spectrum produced by the stellar populations however, are not attenuated using a dust law. 
\par Given the above parameterization of the escape of ionizing photons in the CIGALE SED analysis, our second estimate of the escape fraction, which we denote as $\rm{f}^{\rm{SED}}_{\rm{esc},\rm{LyC}}$ in Table \ref{tab:info} also mimics the definition of the absolute escape fraction. The values of $\rm{f}^{\rm{SED}}_{\rm{esc},\rm{LyC}}$ in Table \ref{tab:info} and their corresponding errors pertain to the likelihood weighted analysis of this quantity over the entire grid of models as defined in Table \ref{tab:cigalegrid}. In CIGALE \citep{Boquien19}, this is presented as the \textit{pdf} analysis.

\begin{figure*}[ht!]
\nolinenumbers
\hspace{-0.25cm}
\includegraphics[width=1.0\textwidth]{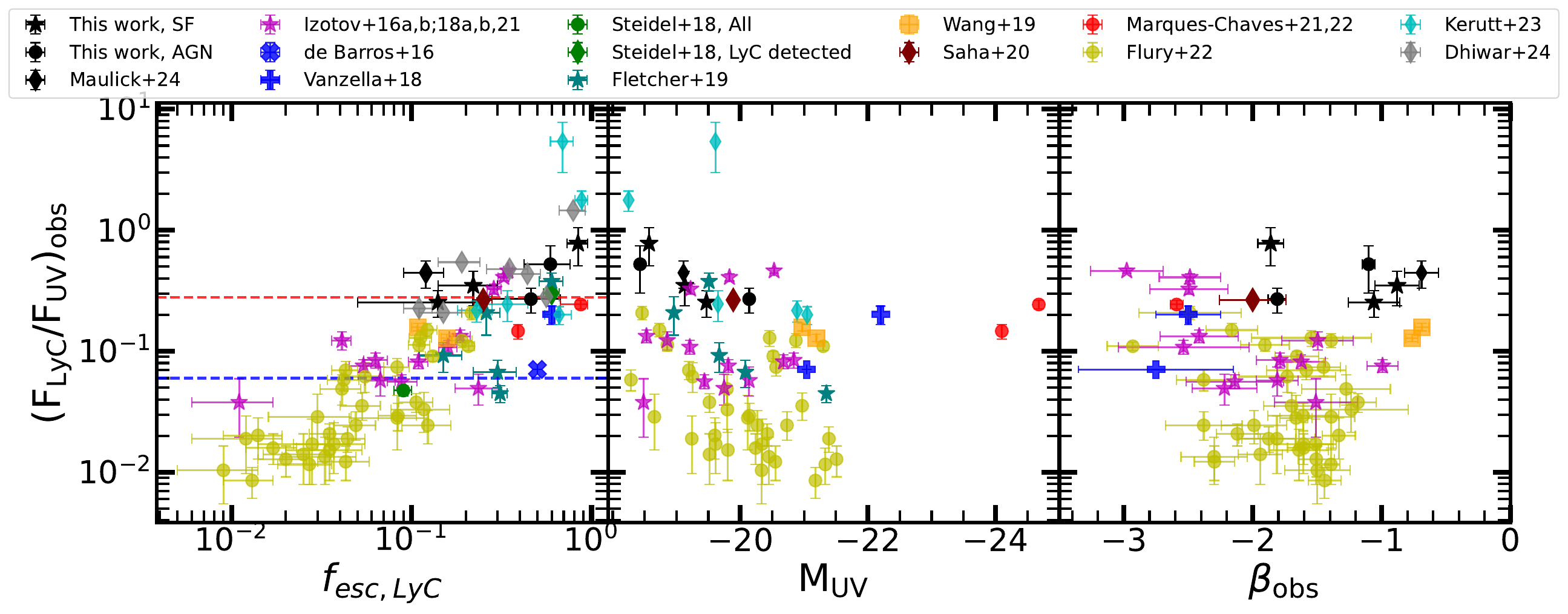}
\caption{In the left panel, we compare $\rm{f}_{\rm{esc,LyC}}$ vs $(\rm{F}_{\rm{LyC}}/\rm{F}_{\rm{UV}})_{\rm{obs}}$ for our sample (star-forming indicated by the black stars and AGN by the black circles) and for those in the literature. Both $\rm{f}_{\rm{esc,LyC}}$ and $(\rm{F}_{\rm{LyC}}/\rm{F}_{\rm{UV}})_{\rm{obs}}$ have been corrected for the IGM attenuation. Individual leakers at low redshift ($z\sim 0.2-0.4$) are from the LzLCS survey \citep{Flury22a}, \cite{Izotov2016,Izotov16b,Izotov18,Izotov18b,Izotov21} and \cite{Wang19}. Intermediate redshift LyC leakers ($z\sim 1.1-1.5$) are from \cite{Saha20}, \cite{Dhiwar24} and \cite{Maulick24}. At high-z ($z>2$) we include detections from \cite{deBarros16, Vanzella16, Vanzella18, Fletcher19,MarquesChaves21,MarquesChaves22} and \cite{Kerutt23}. We also include estimates from the stack of individually detected LyC galaxies and from the stack of the entire sample of \cite{Steidel18}. The red and blue dashed horizontal lines represent the value of $(\rm{F}_{\rm{LyC}}/\rm{F}_{\rm{UV}})_{\rm{obs}}$ assuming a 3$\sigma$ detection in the F154W band and an $\rm{M}_{\rm{UV}}$ equal to -18.50 and -20.14 respectively. In the middle and right panels we compare the observed absolute UV magnitudes ($\rm{M}_{\rm{UV}}$) and the observed UV-continuum slopes ($\beta_{\rm{obs}}$) of our sample to those in the literature.}  
\label{fig:lit_plot}
\end{figure*}

\begin{figure}[ht!]
\nolinenumbers
\includegraphics[width=0.45\textwidth]{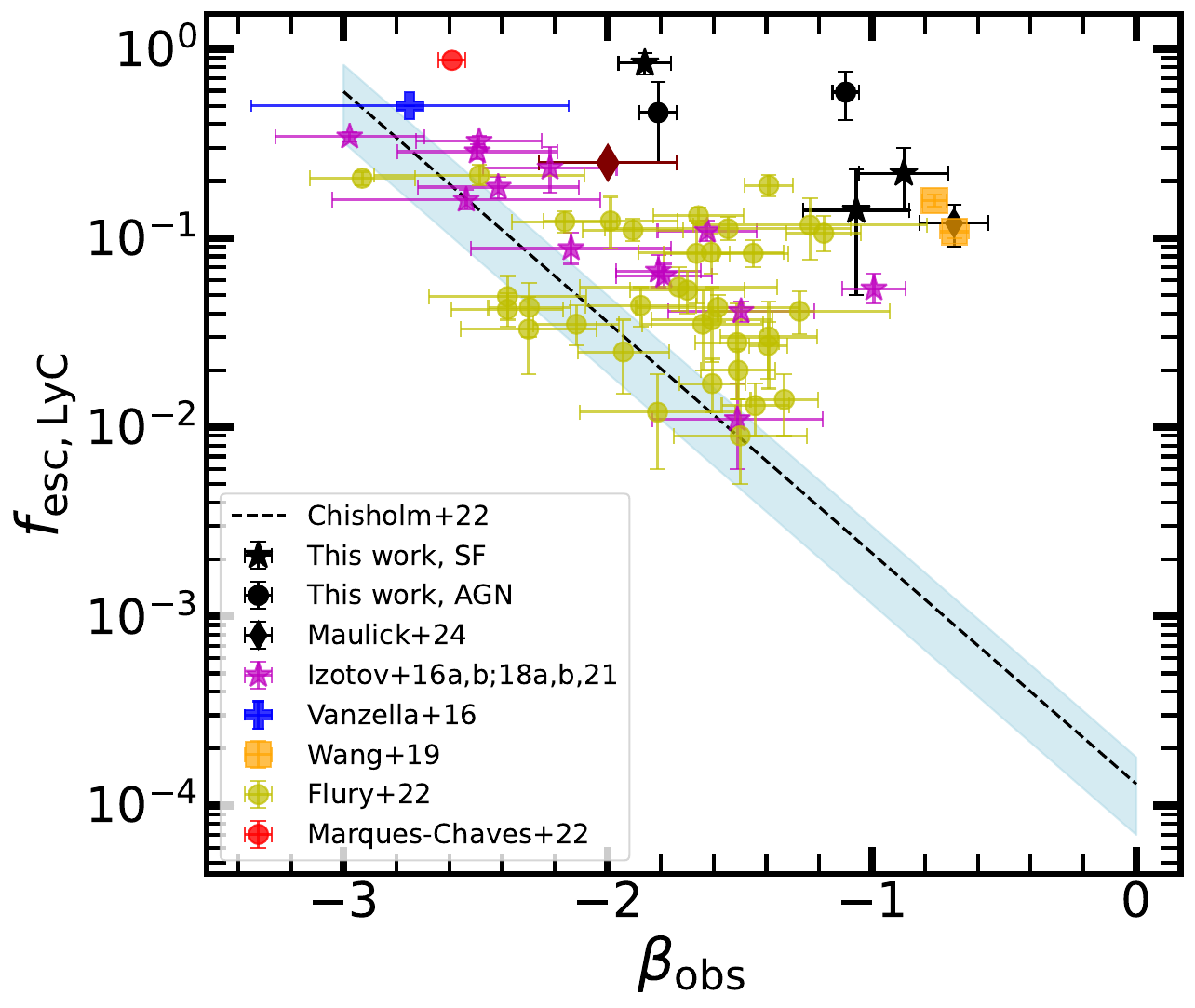}
\caption{ $\beta_{\rm{obs}}$ vs $ \rm{f}_{\rm{esc,LyC}}$ for the galaxies studied in this work and some of the individual LyC leakers from the literature. The empirical relation derived by \cite{Chisholm22} and its 1-$\sigma$ scatter is denoted by the black dashed line and the blue shaded region respectively.} 
\label{fig:chisholm_check}
\end{figure}

\begin{figure}[]
\nolinenumbers
\includegraphics[width=0.5\textwidth]{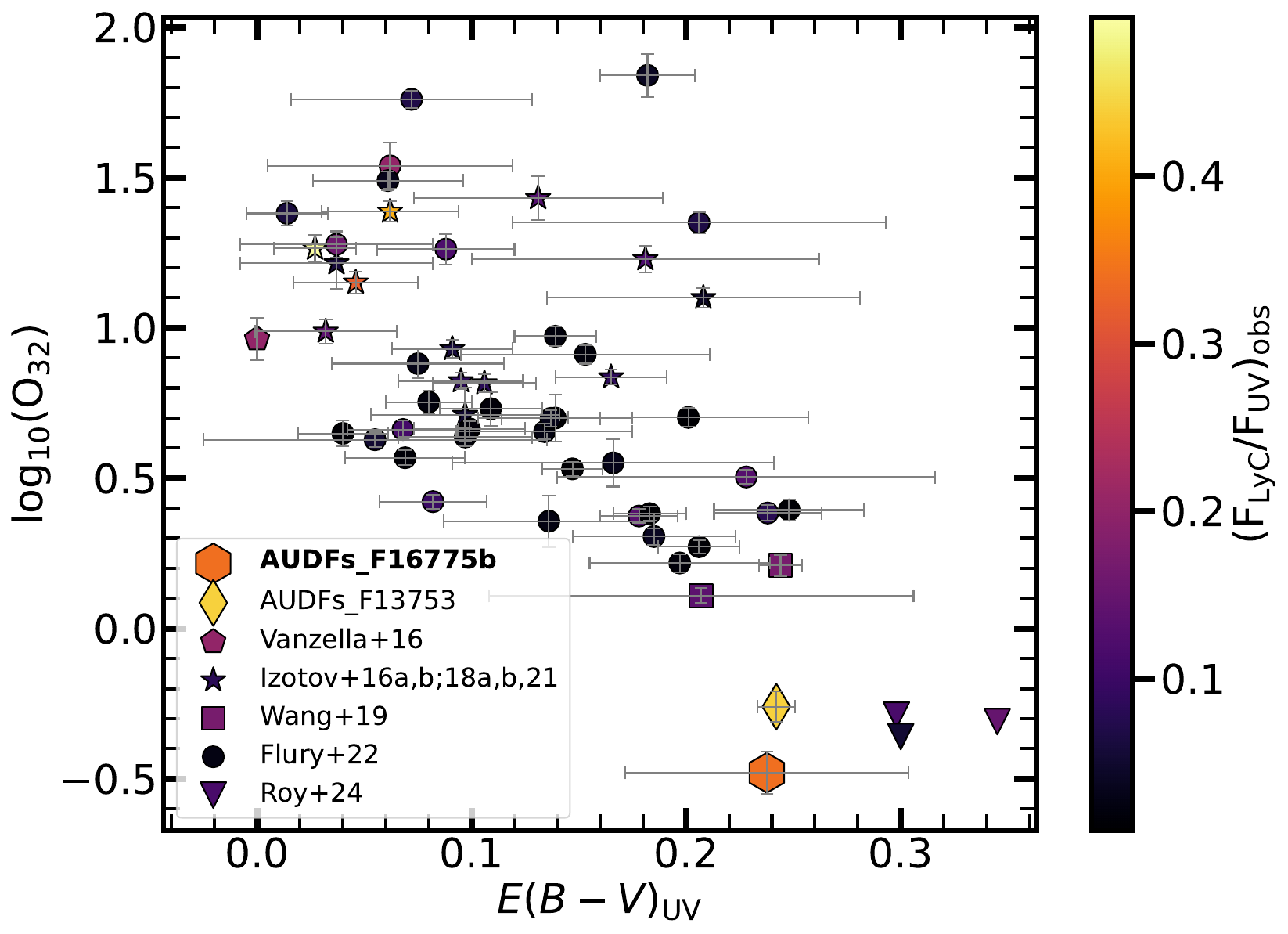}
\caption{The UV color excess $E(B-V)_{\rm{UV}}$ vs $\rm{log}_{10}(O_{32})$ for the galaxies \textbf{AUDFs\_F16775b}, AUDFs\_F13753 (Maulick et al.) and some of the individual LyC leakers from the literature. The points are colored according to their $(\rm{F}_{\rm{LyC}}/\rm{F}_{\rm{UV}})_{\rm{obs}}$ (corrected for IGM attenuation) values indicated by the colorbar on the right.} 
\label{fig:EBV_O32}
\end{figure}

\section{Discussion} \label{sec:discussion}
In this section, we attempt to place our sample in the broader context of other observational LyC studies. 
\subsection{Comparison with other LyC leaker studies}
In Figure \ref{fig:lit_plot} we compare some of the properties of the identified LyC leakers in this work to those in the literature. Namely, we consider individual LyC leakers at low redshifts ($z\sim 0.2-0.4$) from the Low-redshift Lyman Continuum Survey (LzLCS) survey (\cite{Flury22a} and detections from \cite{Izotov2016,Izotov16b,Izotov18,Izotov18b,Izotov21} and \cite{Wang19}). At intermediate redshifts ($z\sim 1-1.5$), we consider LyC leakers from \cite{Saha20}, \cite{Dhiwar24}) and \cite{Maulick24}; and at $z>2$ from \cite{deBarros16}, \cite{Vanzella16,Vanzella18}, \cite{Fletcher19}, \cite{Steidel18}, \cite{MarquesChaves21,MarquesChaves22} and 
 \cite{Kerutt23}. The LyC escape fraction we consider for our sample is the $f^{\rm{H}\alpha}_{\rm{esc,LyC}}$ estimated in Section \ref{sec:fesc_Halpha}. 
  \par In the left panel of Figure \ref{fig:lit_plot}, there appears a significant correlation between the absolute escape fraction and the ionizing to non-ionizing flux density ratio \big($(\rm{F}_{\rm{LyC}}/\rm{F}_{\rm{UV}})_{\rm{obs}}$\big). We correct both these quantities for the IGM attenuation. {The flux density ratio we adopt for this comparison is in frequency space.} For the works of \cite{Vanzella18,Steidel18, Fletcher19,MarquesChaves21,MarquesChaves22,Kerutt23}, this is expected since the definition of the absolute escape fractions adopted in these works is of the form $f_{\rm{esc},\rm{LyC}} \propto (\rm{F}_{\rm{LyC}}/\rm{F}_{\rm{UV}})_{\rm{obs}}$ without a term relating to the dust attenuation at the UV wavelengths (see for example \citep{Steidel18}). In the case of this work, \cite{Saha20}, \cite{Flury22a} (we compare with their $f_{\rm{esc},\rm{LyC}}$ derived from the H$\beta$ recombination line), \cite{Dhiwar24} and \cite{Maulick24}, there is a weaker dependence, since in these works the $f_{\rm{esc},\rm{LyC}}$ is of the form $f_{\rm{esc,LyC}}=\rm{c}_{*}\rm{F}_{\rm{LyC}}/(\rm{c}_{*}\rm{F}_{\rm{LyC}} + \rm{g}{(\rm{H}\beta)})$,
where $\rm{c}_{*}$ is a constant and $\rm{g}_{}$ is a function of either the H$\beta$ or H$\alpha$ dust-corrected luminosity. \par In the following discussion we anchor our comparisons to $(\rm{F}_{\rm{LyC}}/\rm{F}_{\rm{UV}})_{\rm{obs}}$ since this is estimated in a near uniform manner across studies.
However, even when comparing the observed $(\rm{F}_{\rm{LyC}}/\rm{F}_{\rm{UV}})_{\rm{obs}}$, it is important to note the differences in the LyC and non-ionizing UV wavelengths probed across these studies. For instance, the non-ionizing UV wavelength in $(\rm{F}_{\rm{LyC}}/\rm{F}_{\rm{UV}})_{\rm{obs}}$ corresponds to 1100 \text{\AA} in \cite{Flury22a} in contrast to 1500 \text{\AA} used in the rest of the studies including our work. Furthermore, the LyC wavelengths probed by the majority of these studies lie between 850-910 \text{\AA}. For the LyC detections made by UVIT \citep{Saha20,Dhiwar24,Maulick24} at $z\sim1-1.5$, $\lambda_{\rm{LyC}}\sim 650-800 \text{\AA}$. 
\par We also note the differences in the treatment of the IGM within these studies. The effects of these differences may be non-negligible especially at higher redshifts. At the intermediate redshift-range $z\sim 1.2-1.5$, \cite{Dhiwar24} use the statistics of the full sets of 10000 simulated IGM sightlines generated for each of the LyC leakers identified in their work. This leads to a mean transmission of $\rm{T}_{\rm{IGM}}\sim 0.6$, when using the column density prescription of \cite{Inoue14}. However, the fact that these are individually detected in LyC likely skews the probability distribution of IGM sightlines toward those with larger transmission values than the mean. This bias is well outlined and quantified in the work of \cite{Bassett21}. They show that for individually detected LyC galaxies using the average IGM transmission value at that redshift leads to an overestimate of the true $f_{\rm{esc,LyC}}$ and  $(\rm{F}_{\rm{LyC}}/\rm{F}_{\rm{UV}})_{\rm{obs}}$. They also show that this bias increases for surveys with shallower depths. It is in a similar vein, (by constraining $f_{\rm{esc,LyC}}<1$) that some of the literature \citep{Vanzella16,Vanzella18, MarquesChaves21, MarquesChaves22} we consider for individual high redshift LyC leakers ($z>2$) adopt transmission values that are larger than the mean IGM transmission values predicted by models such as those of \cite{Inoue14} at the said redshift.
\par It is interesting that for individually detected LyC galaxies, the region of large $f_{\rm{esc,LyC}}$ and $(\rm{F}_{\rm{LyC}}/\rm{F}_{\rm{UV}})_{\rm{obs}}$ (left panel of Figure \ref{fig:lit_plot}) is almost exclusively occupied by high-redshift galaxies. We suspect that this is a selection effect. While the IGM transmission bias may be one such factor that contributes to this selection effect,  we primarily attribute the large $(\rm{F}_{\rm{LyC}}/\rm{F}_{\rm{UV}})_{\rm{obs}}$ of our candidates to the observational depth of the survey itself. Assuming typical UV bright star-forming galaxies at $z\sim 1$, which is approximately the redshift we probe in this work, corresponding to $\rm{M^{*}_{\rm{UV}}}\sim -18.5$ (adopted from \cite{Bhattacharya23,Sun23}), the lower limit of $(\rm{F}_{\rm{LyC}}/\rm{F}_{\rm{UV}})_{\rm{obs}}$ of LyC candidates that are detected at greater than 3$\sigma$ confidence in the AUDFs F154W band that probes the LyC, is 0.28. We denote this value by the dashed red line in the left panel of Figure \ref{fig:lit_plot}. Adopting an $\rm{M}_{\rm{UV}}$ equal to that of our UV brightest galaxy ($\rm{M^{*}_{\rm{UV}}}\sim -20.1$), leads to a lower limit of 0.06, denoted by the dashed blue line. Thus, our estimates of $(\rm{F}_{\rm{LyC}}/\rm{F}_{\rm{UV}})_{\rm{obs}}$ lie above this value for our entire sample. We note that these lower limits were corrected for the IGM using the same value that was used for estimating $(\rm{F}_{\rm{LyC}}/\rm{F}_{\rm{UV}})_{\rm{obs}}$ and $\rm{f}_{\rm{esc,LyC}}$. The apparent correlation in the left panel of Figure \ref{fig:lit_plot} would then suggest that these observations bias us towards selecting galaxies characterized by relatively high escape fractions. Note \cite{Siana10} probe the LyC emission at wavelengths close to 700 \text{\AA} for 15 galaxies at $z\sim 1.3$. From their stacked non-detection, they infer a LyC to UV flux ratio upper limit of $\sim0.006$. \cite{Alavi20} find an upper limit of 0.009 using a stack of 11 star-forming galaxies between redshifts 1.2 and 1.4.
\par The middle panel of Figure \ref{fig:lit_plot} indicates that for a fixed non-ionizing emitted UV luminosity, the emitted LyC from leakers can take on a wide range of values. The intrinsic ratio, $L_{\rm{LyC,int}}/L_{\rm{UV,int}}$, is shown to be highly dependent on the age, star-formation history (single burst, exponential decay, constant), IMF and the metallicity of the stellar population \citep{Siana07}. In addition to these factors, the emergent ratio (when corrected for IGM attenuation) will depend on the interstellar medium (ISM) properties at the site of LyC production as well as the relative dust attenuation faced by the photons at LyC and UV wavelenghts. This may contribute to the significant scatter in the middle panel of Figure \ref{fig:lit_plot}. We note that the $L_{\rm{LyC,int}}/L_{\rm{UV,int}}$ values corresponding to the \textit{bc03} stellar populations from our best fit CIGALE SEDs for our sample range from $\sim 0.1-0.4$.
\par The UV-continuum spectral slope ($\beta_{\rm{obs}}$) is thought to be linked to LyC photon escape as it depends on the nebular emission generated by ionizing photons in the ISM \citep{Zackrisson13}. Its ease of calculation from photometric measurements makes it an attractive quantity that can be accessed across a range of redshifts. One of the challenges however of analyzing LyC escape through this quantity is its added dependence on stellar population age and reddening \citep{Reddy18}. Empirically, based on the complete LzLCS+ sample (the LzLCS+ sample comprises of leakers from \cite{Flury22a}, \cite{Izotov2016, Izotov16b,Izotov18,Izotov18b,Izotov21} and \cite{Wang19}), including non-detections, \cite{Chisholm22} find a significant correlation between $\beta_{\rm{obs}}$ and $\rm{f}_{\rm{esc,LyC}}$. 
\par For our $\beta_{\rm{obs}}$ comparison (third panel of Figure \ref{fig:lit_plot}), we note that for the sample of \cite{Flury22a}, $\beta_{\rm{obs}}$ is estimated using HST-COS spectra over the range 1050-1350 \text{\AA}, whereas our measurements are carried out in the range 1268-2580 \text{\AA}, following \cite{Calzetti00}. As \cite{Chisholm22} note, the $\beta_{\rm{obs}}$ measurements of \cite{Flury22a} carried out at $\sim 1150 \text{\AA}$ are on average redder by about 14 percent than the values obtained from stellar continuum modeling of the same galaxies at wavelengths $\sim 1500 \text{\AA}$ because of the steepening of dust attenuation curves at bluer wavelengths. The highest $(\rm{F}_{\rm{LyC}}/\rm{F}_{\rm{UV}})_{\rm{obs}}$ occur at the steepest values of $\beta_{\rm{obs}}$ for the LzLCS+ sample.  
 In Figure \ref{fig:chisholm_check}, we highlight the empirical relation between $\beta_{\rm{obs}}$ and $\rm{f}_{\rm{esc,LyC}}$ derived by \cite{Chisholm22} and the location of individually detected LyC galaxies from our sample and from the literature in this plane. Our sample is offset from this relation. 
 We do not emphasize the non-detections in this work though we do note that \cite{Flury22b} and more recently \cite{Jung24}, who studied a sample of blue low-mass lensed galaxies at $z\sim 1.3-3$, both report non-detection of LyC in galaxies with considerably blue $\beta_{\rm{obs}}$ ($\lesssim -2$).
\par AUDFs\_F20788b, the galaxy that has the highest values of  $(\rm{F}_{\rm{LyC}}/\rm{F}_{\rm{UV}})_{\rm{obs}}$ and $f_{\rm{esc,LyC}}$ among our sample, also has the steepest $\beta_{\rm{obs}}$ value (-1.86). Additionally, this galaxy is low-mass ($< 10^{9} \rm{M}_{\odot}$) and has a Balmer decrement that is consistent with negligible dust content. \textbf{AUDFs\_F21146}, \textbf{AUDFs\_F16775b}, and AUDFs\_F13753 (from \cite{Maulick24}) on the other hand have moderate $\beta_{\rm{obs}}$ values ($\gtrsim -1$) and have relatively larger stellar masses ($>10^{9.8} \rm{M}_{\odot}$). The nebular reddening $\rm{E(B-V)_{neb}}$, for these galaxies are also modest $\gtrsim 0.5$. These galaxies have the lowest $(\rm{F}_{\rm{LyC}}/\rm{F}_{\rm{UV}})_{\rm{obs}}$ and $\rm{f}_{\rm{esc,LyC}}$ among our sample. Interestingly, all three of these galaxies have blue clumpy star-forming regions (Section \ref{sec:offset_lyc}). In \cite{Maulick24} we report AUDFs\_F13753 is a likely major merger based on high resolution JWST NIRCam imaging \citep{Rieke23}. We discuss the morphology of LyC leaking galaxies further below.
 \subsection{The diverse morphology and properties of LyC leakers}
The morphology of LyC leakers across different bands may provide insights into the nature of LyC escape from these galaxies. In the local universe, HST imaging of the LyC leaker Haro 11 \citep{Bergvall06} reveals a disturbed structure comprised of star-forming knots that display signatures of dwarf galaxy mergers \citep{Komarova24}. The [O III] $\lambda$ 5007, H$\alpha$ and [O I] $\lambda$ 6300 imaging of Haro 11 \citep{Menacho19} using MUSE reveals potential links between the stellar and mechanical feedback, and the LyC leakage from the system. The LzLCs sample \citep{Flury22a} is comprised of extremely compact galaxies of the 'Green Peas' type, with most of the detected leakers in the sample having UV half-light radii less than 1 kpc. High-resolution HST imaging of several Green Pea galaxies has revealed clumpy morphologies and low surface-density features, indicating ongoing mergers \citep{Cardamone09}.   
\par In Section \ref{sec:offset_lyc} we highlighted some of the galaxies in our sample that have an extended morphology and in which we observe blue UV clumps. Clumpy UV structures are not uncommon in high redshift star-forming galaxies \citep{Elmegreen05, Law07, Guo15}. 
There exist multiple pathways for the formation of these clumps, with gravitational disk instabilities, minor and major mergers all thought to play a role \citep{Elmegreen05,Guo15,Calabro19}. The LyC emission from these systems in our work however, seem to be offset from the stellar FUV continuum emission. 
So far, a firm theoretical understanding of spatially offset Lyman-continuum emission has not been established. At similar redshifts to that probed in this work, \cite{Dhiwar24} report offsets of $\sim$0.8", similar to what we find in this work, corresponding to $\sim 5-6$ kpc. However, like in this work the LyC emission was detected in the UVIT F154W band which has a coarser PSF  (FWHM $\sim$ 1.6") than the HST. At higher redshifts ($z\sim 3-4$), studies using the HST F336W band to probe LyC emission have found offsets between the LyC and the non-ionizing UV emission (probed by the optical and IR HST bands). At $z=3.14$, \cite{Mostardi15} report the detection of LyC in the HST F336W band from a galaxy that is comprised of two clumps in the observed optical and IR bands. The LyC emission was detected only in one of the two clumps. \cite{Yuan24,Yuan21} ($z=3.80$) and \cite{Gupta24} ($z=3.09$) report LyC emission from galaxies in which the LyC centroid is offset from rest-frame NUV center of the galaxy by about 0.2$^{''}$ (1.4 kpc) and 0.3$^{''}$ (2.2 kpc) respectively. Similar offsets have also been reported by \cite{Micheva17}. Compiling LyC leaking candidates at $z>3$ from the literature, \cite{Yuan24} report that 60-70 $\%$ of the candidates show spatially offset LyC emission ($\gtrsim 1.4$ kpc) from the rest-frame non-ionizing UV continuum. These include 18/29 detections in which the LyC is traced by high resolution HST imaging (F336W).
\par Given the large integrated dust content inferred from the Balmer decrement and the 'red' nature of the central regions of galaxies \textbf{AUDFs\_F21146} and \textbf{AUDFs\_F16775b}, it is more likely that these relatively 'blue' regions in these extended systems are in someway associated with the sites of the escaping LyC emission. The UV dust attenuation has been found to be empirically related to the neutral gas covering fraction \citep{Reddy16}. Thus the geometry in the ISM that allows the escape of LyC photons from these systems presents a challenge. 
\par At low redshifts ($z\sim 0.2-0.4$), the majority of the identified LyC leakers are low-mass, compact, have negligible UV dust attenuation and large values of the $[\text{O III}] \lambda 5007/ [\text{O II}] \lambda 3727$  (O32) ratio ($>3$) \citep{Izotov2016,Izotov18,Izotov21,Flury22b, SaldanaLopez22}. High O32 values are thought to be a necessary but insufficient condition for LyC leakage. Another indirect spectroscopic indicator of LyC leakage proposed by \cite{Wang19} is the weakness of [S II] 6717, 6731 emission lines. A high O32 and relatively weak [S II] emission are both thought to be associated with density-bounded LyC escape. Weak emission of [O II] and [S II] that originate from the outer neutral edges of the Str{\"o}mgren sphere would imply that this outer region is depleted. We display the curve used by \cite{Wang19} to quantify the deficiency of [S II] emission in Figure \ref{fig:bpt_sii}. The [S II] deficiency is defined as the displacement in $\rm{log}(\rm{[S II]/\rm{H}\alpha})$ from this curve. Only \textbf{AUDFs\_F16775b} from our sample has a significant [S II] deficiency comparable to the two leakers identified in \cite{Wang19} ($\sim -0.30$ dex).
\par For \textbf{AUDFs\_F16775b} and AUDFs\_F13753 (from \cite{Maulick24}), available MUSE spectra from the MUSE Hubble Deep Field Surveys DR2 \citep{Bacon23} allows us to estimate the extinction corrected O32 values of these two galaxies, with the [O II] $\lambda 3727$ detected in the MUSE spectra. We highlight the extinction corrected O32 vs UV color excess of these galaxies and for some of the individual LyC leakers from the literature in Figure \ref{fig:EBV_O32}. We estimate $E(B-V)_{\rm{UV}}$ for \textbf{AUDFs\_F16775b} and AUDFs\_F13753 by converting the $E(B-V)_{\rm{neb}}$ obtained from the Balmer decrement, using the relation $E(B-V)_{\rm{UV}}=0.44\times E(B-V)_{\rm{neb}}$ \citep{Calzetti00}. The relation $E(B-V)_{\rm{UV}}=E(B-V)_{\rm{neb}}$, occasionally used in high redshift studies (e.g., \cite{Bassett19}) will result in a larger UV dust attenuation and hence will not alter the inferences derived from the positions of these two galaxies in this plane relative to the literature sample. For the LzLCS+ sample, we adopt the $E(B-V)_{\rm{UV}}$ values from \cite{SaldanaLopez22}. The low-O32 values for \textbf{AUDFs\_F16775b} and AUDFs\_F13753 ($M_{*}\sim 10^{10.32} M_{\odot}$) are consistent with the findings of \cite{Paalvast18} who studied the relation between O32 and the stellar mass for a sample of galaxies between redshifts 0.28 and 0.85. While they do not find a statistically significant anticorrelation, it is worth noting that the majority of their sample with stellar masses $\rm{log}_{10}(M_{*}/M_{\odot})>9$ have $\rm{log}(O32)$ values $<0$.
\par The two [S II] deficient LyC leakers (J0910 and J1432) identified in \cite{Wang19} are similar to \textbf{AUDFs\_F16775b} and AUDFs\_F13753 in that, they too are characterized by modest UV dust attenuation ($\gtrsim 0.2$), large stellar masses ($\rm{log}_{10}(M_{*}/M_{\odot})>10.4$) and relatively low O32 values (1.29 and 1.57 respectively). More directly, the $E(B-V)_{\rm{neb}}$ (0.46 and 0.28 respectively) derived from the Balmer decrement and $\beta_{\rm{obs}}$ (-0.69 and -0.76 respectively) for these two leakers are comparable to the values for \textbf{AUDFs\_F21146}, \textbf{AUDFs\_F16775b} and AUDFs\_F13753. J0910 and J1432 also have similar $\rm{SFR}_{\rm{H}\alpha}$ (35 and 19 $M_{\odot} \rm{yr}^{-1}$ respectively) and relatively weak rest-frame H$\alpha$ equivalent widths (138 and 113 \text{\AA} respectively). The morphologies of J0910 and J1432 however are similarly compact to other Green Pea leakers at $z\sim 0.2-0.4$ (half light radii $<0.6$ kpc).  In Figure \ref{fig:chisholm_check} these five galaxies have the largest offsets in the $\beta_{\rm{obs}}$ vs $\rm{f}_{\rm{esc,LyC}}$ relation. {This offset may indicate a highly inhomogeneous ISM and/or a clumpy dust geometry, where LyC photons escape through dust-free channels with small opening angles. Hints of a similar complex geometry are observed in Haro 11, where Knot B—the strongest LyC emitter \citep{Komarova24} among its three star-forming knots (A, B, and C)—exhibits the shallowest UV continuum slope ($\beta_{\text{obs}}=-0.86$) while also having the highest nebular dust content ($E(B-V)_{\rm{neb}}=0.38$) \citep{Ostlin21}. Recently, \cite{Roy24} identified a sample of three new LyC leakers selected for their high stellar mass and [S II] deficiency, exhibiting properties similar to the two LyC leakers in \cite{Wang19}. These leakers also have low O32 values and are positioned to the right of our points in Figure \ref{fig:EBV_O32}.} Thus, \textbf{AUDFs\_F21146}, \textbf{AUDFs\_F16775b} and AUDFs\_F13753, together with the two identified leakers in \cite{Wang19} and {the three newly detected leakers in \cite{Roy24}} represent some of the most massive dust obscured LyC leakers with low O32 values (also see \citep{Bassett19}).
We thus believe that it is worth reassessing the importance of these massive dusty LyC leakers to calculations that use indirect estimators like the $\beta_{\rm{obs}}$ in estimating the ionizing photon budget (e.g., \cite{Chisholm22}) in the EOR. Updating the statistics on the fraction of LyC leakers among galaxies with similar properties, at different redshifts, would be a first step towards the above. We emphasize however, that while the integrated properties of these galaxies may at face value make these galaxies unfavorable candidates for LyC leakage, the blue star-forming regions within these galaxies are likely sites of significant intrinsic LyC photon production. Spatially resolved measurements, such as that of $\beta_{\rm{obs}}$, may reveal these substructures to be similar to known LyC leakers. However, such an analysis is outside the scope of this work. 

\section{Summary}\label{sec:summary}
\par We summarize the findings of this work below:
\begin{enumerate}
    \item Using archival data from the HST (CLEAR, \cite{Simons23}), JWST (JADES NIRSpec, \cite{Bunker23}) and the VLT/MUSE (MUSE HUDF, \cite{Bacon23}) in conjunction with UVIT FUV and NUV imaging of the GOODS South (AUDF South, \cite{Saha24}), we report the detection of 5 new LyC leaker candidates in the redshift range $\sim 1-1.5$. 
    \item High-resolution HST UV imaging reveals that some of these leaker candidates have irregular morphologies with extended features and star-forming clumps. We report tentative evidence of the LyC emission from this subset of galaxies in our sample being spatially offset from the non-ionizing UV continuum.
    Two of the five identified LyC leakers have line ratios consistent with AGN activity. The SED-derived stellar masses and the observed UV luminosities of the galaxies studied in this work range from $\rm{log}_{10}(M_{*}/M_{\odot})=8.74-9.85$ and $\rm{M}_{UV}=-18.57\:\rm{to}\:-20.14$ respectively.
\item The observed LyC to non-ionizing flux ratio, after correcting for IGM attenuation, and the LyC escape fraction of these galaxies rank them among the most prominent LyC leakers identified in the literature. However, we propose that this prominence could be influenced by selection effects, which may also affect other high-redshift LyC leaker surveys. 
\item A subset of these LyC leakers have characteristics typically considered unfavorable for LyC leakage (large stellar mass, red UV continuum slopes and low O32). While we find a couple of examples of similar leakers in the literature, this discrepancy calls for further investigation into the underlying physics of LyC leakage from these systems and possible statistical importance of these systems to reionization.
\end{enumerate}
Some of the data presented in this article were obtained from the Mikulski Archive for Space Telescopes (MAST) at the Space Telescope Science Institute. The specific observations analyzed can be accessed via the following \dataset[DOI]{http://dx.doi.org/10.17909/84cz-gm21}. \\
We thank the anonymous referee for their careful review and suggestions that have improved the quality of this manuscript.

\facilities{AstroSat (UVIT), HST (WFC3, ACS-WFC, WFC3-UVIS), JWST (NIRSpec), VLT (MUSE)}


\software{astropy \citep{2013A&A...558A..33A,2018AJ....156..123A},  
          CIGALE \citep{Boquien19}, pypher \citep{Boucaud2016}
          }


\appendix
\section{Spectroscopic data}
\label{sec:1d_spec}
\begin{figure*}[ht!]
\nolinenumbers
\includegraphics[width=1\textwidth]
{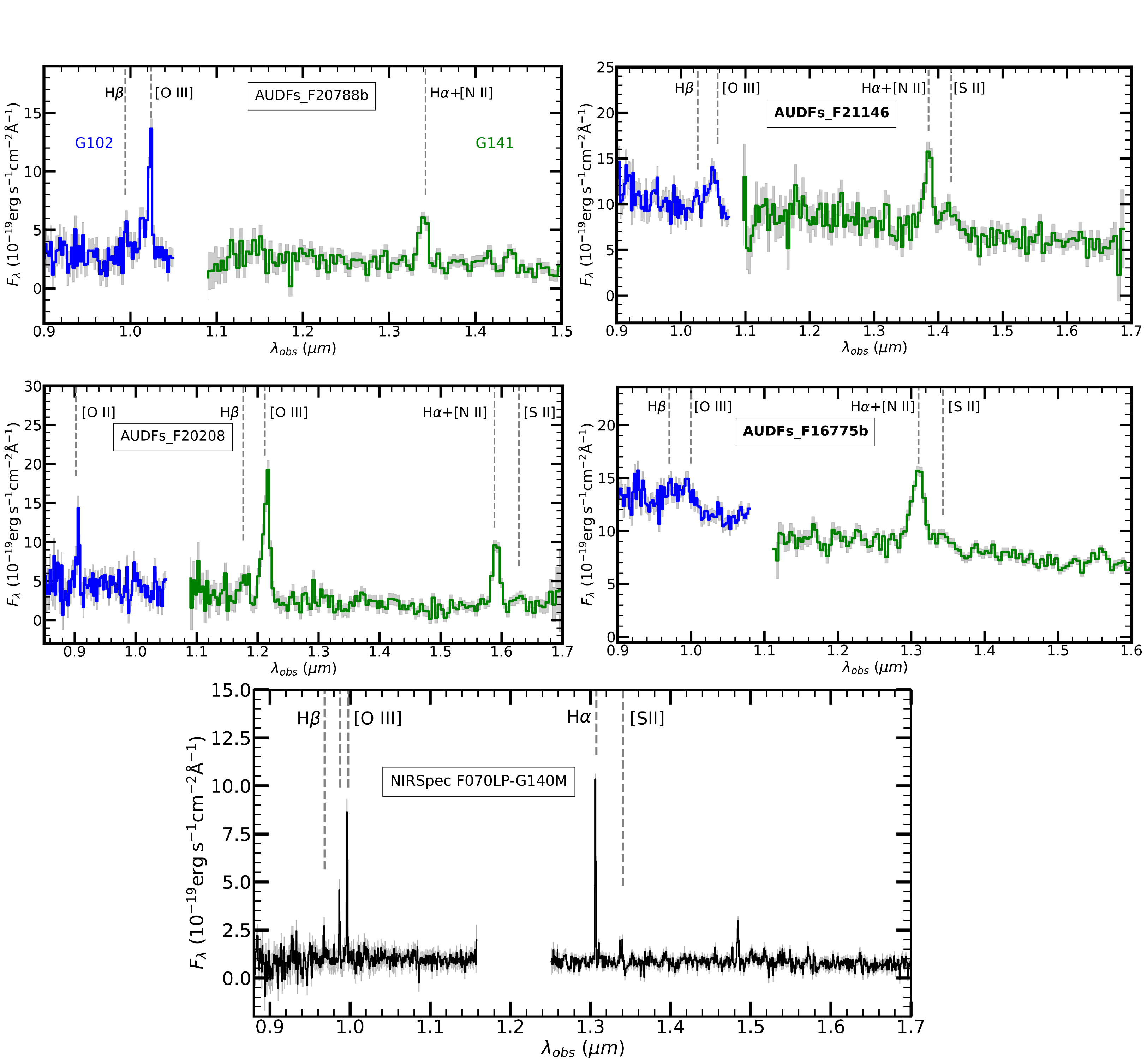} 
\caption{Science grade 1D spectra from publicly available data releases of the CLEAR \citep{Simons23} and JADES NIRSpec \citep{Bunker23} surveys  covering the rest-frame optical emission lines for the galaxies in our sample. For the CLEAR spectra, the blue curves represents the HST G102 grism spectra while the green curves represent the HST G141 grism spectra. The grey curves in all the plots represent the error spectra. The vertical dashed lines indicate the emission lines in the spectra as reported by \cite{Simons23}.}

\label{fig:optical_spec}
\end{figure*}

In Figure \ref{fig:optical_spec} we highlight the 1D spectra of the galaxies in our sample containing major rest-frame optical emission lines like H$\alpha$ and the [O III] $\lambda \lambda$ 4960,5007 doublet. For 4 of the 5 galaxies, we display the science-grade 1D spectra from the CLEAR survey data release \citep{Simons23}. Note that the [O III] $\lambda \lambda$ 4960,5007 doublet is unresolved in the CLEAR spectra. For \textbf{AUDFs\_F14054} we additionally obtain its 1D spectrum from the JADES \citep{Eisenstein23} NIRSpec Initial Data Release for the Hubble Ultra Deep Field \citep{Bunker23} and highlight this spectrum in Figure \ref{fig:optical_spec}. We inspect the 1D spectra to confirm the spectroscopic redshifts of the galaxies in our sample.
\par
\begin{figure*}[ht!]
\nolinenumbers
\includegraphics[width=1\textwidth]
{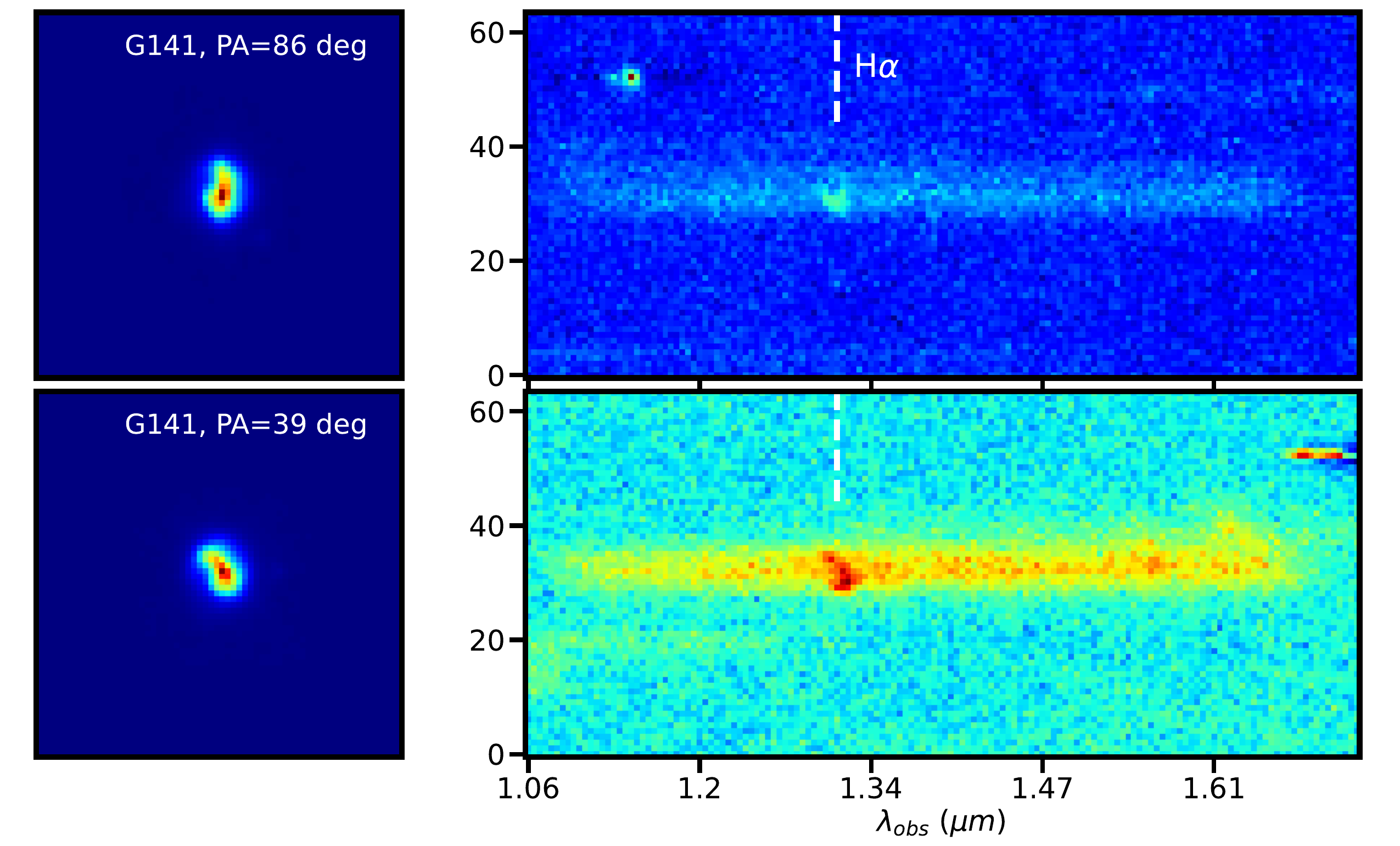} 
\caption{The 2D grism G141 spectra (right) of \textbf{AUDFs\_F16775b} obtained from the CLEAR survey \citep{Simons23}. The direct science image is displayed on the left and the object's orientation is with respect to the position angle of that particular grism pointing. The dashed white line indicates the H$\alpha$ emission line.}

\label{fig:G141_galaxy5_2d}
\end{figure*}

\begin{figure*}[ht!]
\nolinenumbers
\includegraphics[width=1\textwidth]
{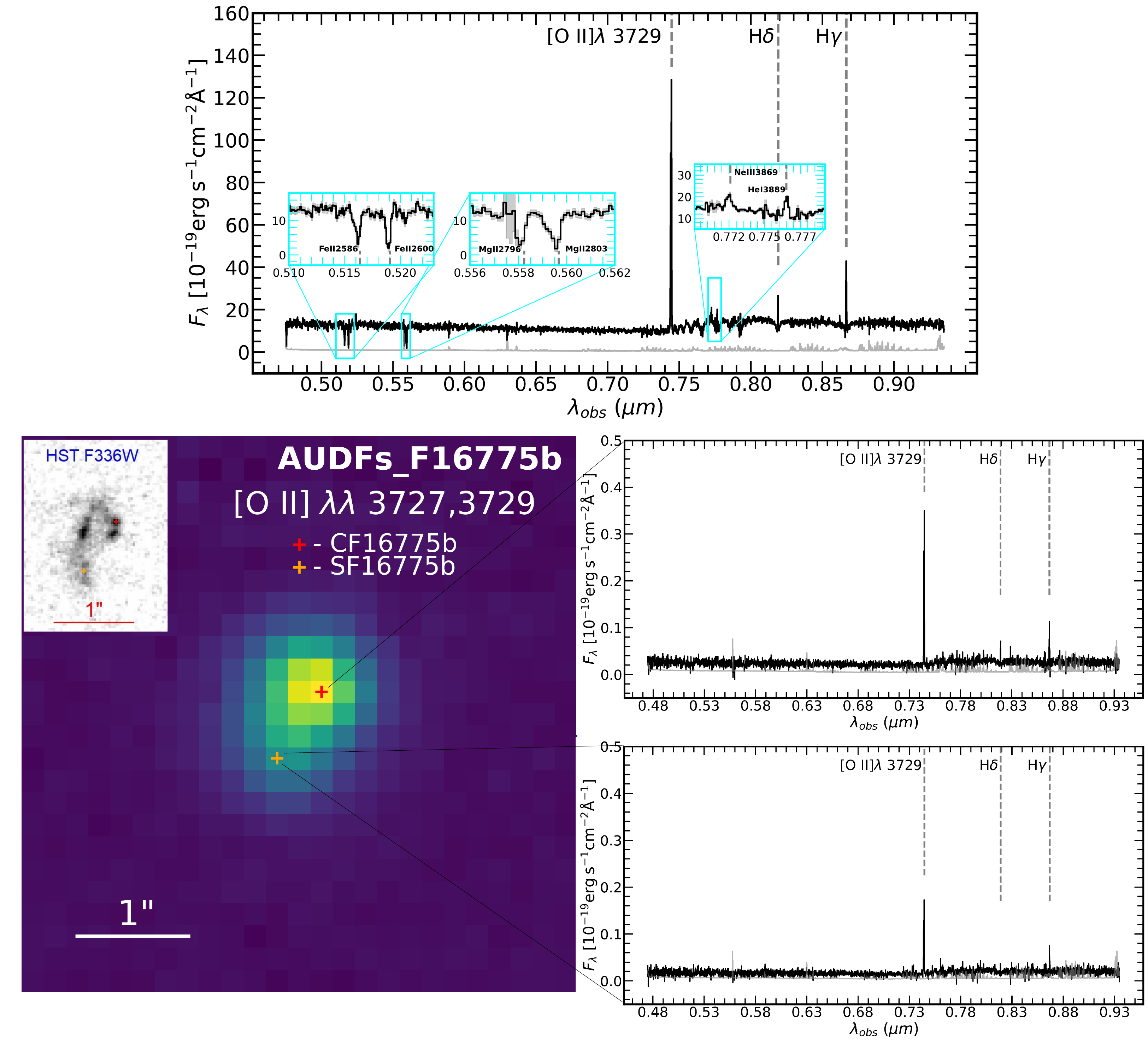} 
\caption{The integrated MUSE 1D spectrum (MUSE HUDF DR2, \cite{Bacon23}) of \textbf{AUDFs\_F16775b} is displayed on the top, with emission and absorption lines in the spectrum marked with the grey dashed lines. The grey curve corresponds to the error spectrum. In the bottom, the MUSE [O II]$\lambda \lambda$ 3727,3729 narrowband image of the object is displayed along with an inset of the HST F336W band image of the object for reference to highlight the differences in the spatial resolution of the two instruments. Representative 1D MUSE spectra extracted from the individual spaxels corresponding to the regions marked by the red and orange crosses are displayed on the right.}

\label{fig:MUSE_galaxy5}
\end{figure*}

 In Figures \ref{fig:G141_galaxy5_2d} and \ref{fig:MUSE_galaxy5}, we present additional plots related to the HST grism and MUSE HUDF data \citep{Bacon23}, respectively, for the galaxy \textbf{AUDFs\_F16775b}. The concern for this galaxy is the potential presence of a foreground interloper, as one of the clumps (CF16775b) appears to be at a photometric redshift of 0.72 (discussed below in Appendix \ref{sec:eazy}). In Figure \ref{fig:G141_galaxy5_2d}, the 2D G141 grism spectra of this object corresponding to the two individually taken position angles of the slitless exposure reveals that the shape of the direct image is reflected in the H$\alpha$ emission. 
\par Assuming that there indeed is a foreground interloper at a redshift of $\lesssim0.72$ in the system, the sensitive MUSE data ($3 \sigma$ line flux limit $\sim 3.1 \times 10^{-19} \text{erg}\: \text{s}^{-1}\text{cm}^{-2}$, \cite{Bacon23}) would likely contain the [O II]$\lambda \lambda$ 3727,3729, the H$\beta$ and the [O III] $\lambda \lambda$ 4959,5007 line complexes corresponding to this redshift. In the integrated spectrum of the object (top row, Figure \ref{fig:MUSE_galaxy5}) however, we report only emission and absorption lines consistent with the spectroscopic redshift of $\sim 1$. We also highlight the narrwoband [O II]$\lambda \lambda$ 3727,3729 image (bottom row, Figure \ref{fig:MUSE_galaxy5}) constructed from the MUSE IFU datacube and representative spectra from 2 individual spaxels of the object. It is worth noting that the spatial resolution of the MUSE observations (FWHM $\sim 0.5"$ at 8000 \AA, \cite{Bacon23}) means that the clumps in the western parts of the system remain unresolved in the MUSE cube data.

\label{sec:galaxy5_spec}

\section{Clump photometric redshift estimation} \label{sec:eazy}
\begin{figure*}[ht!]
\nolinenumbers
\includegraphics[width=1\textwidth]
{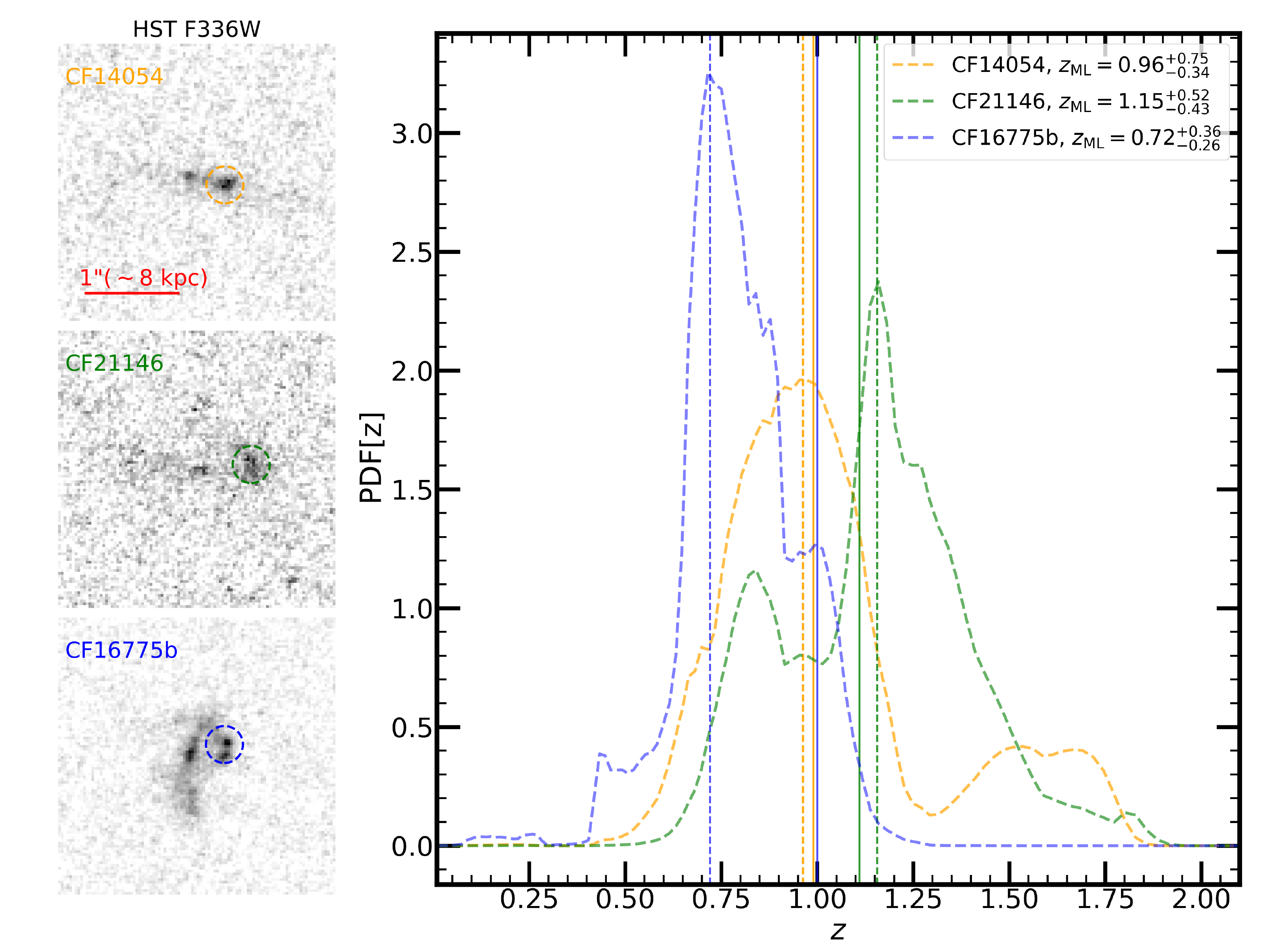} 
\caption{On the left, the location of the clumps are highlighted within the colored circular apertures of radius 0.2" in the HST F336W band. The phot-$z$ PDFs of the clumps are indicated by the dashed curves and the corresponding maximum likelihood values of the phot-$z$ are indicated by the vertical dashed lines on the plot in the right. The values of the maximum likelihood phot-$z$ values are included in the legend. The solid vertical lines represent the spectroscopic redshifts.}

\label{fig:eazy_plot}
\end{figure*}

We attempt to derive the photometric redshifts (phot-$z$) of the clumps we observe in the HST F336W band for the galaxies \textbf{AUDFs\_F14054}, \textbf{AUDFs\_F21146} and \textbf{AUDFs\_F16775b} in this Appendix. We label these clumps CF14054, CF21146 and CF16775b respectively and highlight their locations within apertures of radius 0.2" in Figure \ref{fig:eazy_plot}. For the phot-$z$ estimation, we use these circular apertures with radii approximately matching the PSF FWHM of the HST bands, to compute the flux, thus minimizing light contamination from other regions that could bias our estimates. We compute the fluxes of the clumps in the following HST bands: F336W, F435W, F606W, F775W, F814W, F850LP, F105W and F125W, and use the code \texttt{EAZYpy} \citep{Brammer08,Brammer21} to derive the probability density functions (PDFs) of the phot-$z$. The template spectra we adopt are the PEGASE2.0 template set given its inclusion of a large set of models with varying star-formation histories, ages and dust content. We highlight the PDFs of the clumps in Figure \ref{fig:eazy_plot}. The maximum likelihood phot-$z$ (indicated by the dotted vertical line) of CF14054 and CF21146 nearly match the spectroscopic redshifts (indicated by the solid vertical line). The maximum likelihood phot-$z$ of CF16775b is lower than the spectroscopic redshift though we do notice a peak in the PDF at the spectroscopic redshift.

\section{Veracity of the Source Extractor detections} \label{sec:source_extractor}
{In \cite{Saha24}, we used the narrowest Gaussian filter available in the default Source Extractor filter list to detect sources in the F154W band. This choice was motivated by the aim of detecting faint sources, albeit with the trade-off of potentially identifying noise peaks. Although the minarea parameter, which specifies the minimum number of connected pixels required for source detection, was set to a relatively high value (11 pixels), we found that the convolution filter played a more significant role in suppressing noise peaks. However, this approach may inadvertently prevent the detection of some faint sources. The experiments supporting this conclusion are detailed in Appendix A and C of \cite{Saha24}. In Appendix C of \cite{Saha24} we demonstrated that the probability of background noise fluctuations aligning by chance to mimic objects that are both detected by Source Extractor with our configuration and are brighter than the 3$\sigma$ limit is small. The expected density of such sources is $2\times10^{-4}$ arcsec$^{-2}$}. \par
\begin{figure*}[ht!]
\nolinenumbers
\centering
\includegraphics[width=0.65\textwidth]
{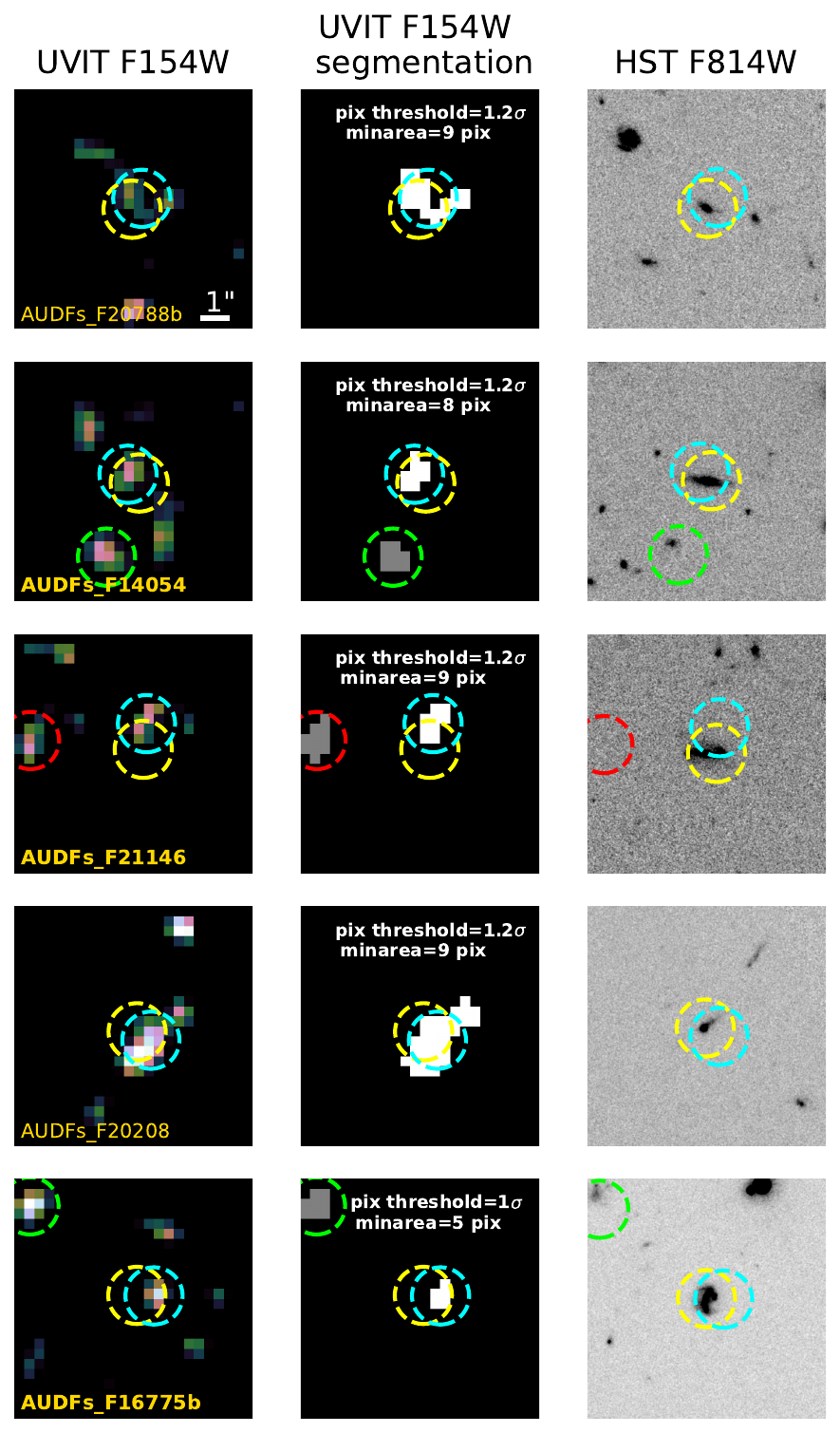} 
\caption{{\textit{Left column}: F154W band images of  $10^{''}\times 10^{''}$ fields centered on the candidate LyC leakers. The image has been convolved with a Gaussian kernel with an FWHM nearly equal the FWHM of the F154W PSF. \textit{Middle column}: The segmentation of sources detected by Source Extractor in the field after convolution with the aforementioned Gaussian kernel. The threshold and minarea parameters used for detection are mentioned in the top of the image. \textit{Right column}: The HST F814W band image of the same field. The cyan apertures are centered on the F154W centroids of the LyC leaking candidates whereas the yellow apertures are centered on the F336W centroids (Figure \ref{fig:stamp_images}). The green apertures enclose neighbouring sources detected in the F154W band for which we are able to identify an HST counterpart within a PSF FWHM distance. The red aperture (field corresponding to AUDFs\_F21146) encloses an object for which are not able to identify any possible HST counterpart. All apertures are of radius  $1.2^{''}$. }}

\label{fig:seg_plot}
\end{figure*}

 { Here, we test the robustness of the LyC-leaking candidate detections by applying a broader Gaussian filter, with an FWHM twice that of the one used in the original detection by \cite{Saha24}. The convolution of the image with this broader filter should result in a more effective suppression of noise peaks. After a combination of different thresholds and minareas, we adopt a threshold of 1.2$\sigma$ and a minarea of 8-9 pixels for this exercise. For reference, the value of the threshold we adopted for the F154W band was 0.8$\sigma$ in \cite{Saha24} and the PSF FWHM of the F154W band is roughly 3 pixels. We then run Source Extractor on $10^{''}\times 10^{''}$ fields of the F154W band centered on our LyC leaking candidates after convolution with a Gaussian kernel with an FWHM of 3 pixels. In the middle panel of Figure \ref{fig:seg_plot} we highlight the segmentation of objects detected with this configuration in the $10^{''}\times 10^{''}$ fields. Except for {AUDFs\_F16775b}, for whose detection we had to lower the pixel threshold to $1\sigma$ and minarea parameter to 5 pixels, the rest of our sources are detected with our fiducial configuration. We highlight the detected LyC candidates in the field, by marking them within cyan and yellow apertures of radius $ 1.2^{''}$, whose centers correspond to the F154W centroid and the F336W centroids respectively.
In three of the five fields, we detect one additional source other than our LyC leaking candidate.  In the cases we are able to identify a possible counterpart in the redder HST band (right panel of Figure \ref{fig:seg_plot}), we highlight this additional source by marking it within a green aperture. Only in the field corresponding to AUDFs\_F21146, we find a detected source in the F154W band for which are we not able to identify a possible HST counterpart. We highlight this source within a red aperture. As alluded to in Section \ref{sec:LyC_sig}, we can use such sources to define false positives: Sources detected by Source Extractor in the F154W band for which we are not able to identify an HST counterpart. We can translate this into a false positive density for the field, which after combining the area of the five $10^{''}\times 10^{''}$ fields corresponds to $1/500=0.002$ arcsec$^{-2}$. }
\par {Since we probe a small area, we also run Source Extractor with the above configuration on a much larger $1{'}\times 1{'}$ single patch of sky in the F154W band.  We detect 82 sources in this bigger field. We then cross match these sources to look for possible HST counterparts using the catalog of \cite{Whitaker19} and identify 9 F154W sources that do not have any HST object 1.6" away from their center. This leads to a false positive density of $9/3600\sim 0.002$ arcsec$^{-2}$. This is in excellent agreement with the one we estimate from the combination of the five smaller $10^{''}\times 10^{''}$ patches.}

\bibliography{AUDFs_lyc_sample_revised_arxiv}{}
\bibliographystyle{aasjournal}



\end{document}